\documentclass[a4paper,11pt]{article}
\pdfoutput=1 

\usepackage[no-natbib-sort]{jheppub}

\usepackage[T1]{fontenc} 

\usepackage{mathtools,amsfonts,amssymb,amsthm}
\usepackage[utf8]{inputenc}
\usepackage{todonotes}
\usepackage{bm}
\usepackage{booktabs}
\usepackage[font=small]{caption}
\usepackage{subcaption}
\usepackage{multirow}
\usepackage{titlesec}
\usepackage{tikz}
\usepackage[centertableaux,boxsize=1.1em]{ytableau}
\usepackage{hyperref}
\usepackage[nameinlink,capitalize]{cleveref}

\makeatletter{}
\g@addto@macro\bfseries{\boldmath}
\renewcommand{\@email}[1]{\texttt{#1}}
\makeatother

\titleformat{\paragraph}[hang]{\normalfont\normalsize\bfseries}{}{0em}{}
\titlespacing{\paragraph}{0pt}{3.25ex plus 1ex minus .2ex}{0.5em}
\crefname{paragraph}{Section}{Sections}

\newcommand*{\seccoord}[1]{\ensuremath\hat{#1}} 
\newcommand*{\secx}{\ensuremath\seccoord{x}} 
\newcommand*{\secy}{\ensuremath\seccoord{y}} 
\newcommand*{\secz}{\ensuremath\seccoord{z}} 
\newcommand*{\ratsec}[1]{\ensuremath\hat{#1}}

\newcommand*{\height}[0]{\ensuremath \tilde{b}}

\DeclareMathOperator{\SO}{SO}
\DeclareMathOperator{\Sp}{Sp}

\DeclareMathOperator{\SU}{SU}
\DeclareMathOperator{\U}{U}
\DeclareMathOperator{\gE}{E}

\DeclareMathOperator{\gG}{G}

\newcommand*{\asu}{\ensuremath\mathfrak{su}}
\newcommand*{\au}{\ensuremath\mathfrak{u}}
\newcommand*{\aso}{\ensuremath\mathfrak{so}}
\newcommand*{\asp}{\ensuremath\mathfrak{sp}}

\newcommand*{\aF}{\ensuremath\mathfrak{f}}
\newcommand*{\aG}{\ensuremath\mathfrak{g}}
\newcommand*{\cN}{\mathcal{N}}
\newcommand*{\cO}{\mathcal{O}}

\newcommand*{\cS}{\mathcal{S}}
\newcommand*{\F}{\mathbb{F}}
\newcommand*{\bP}{\mathbb{P}}
\newcommand*{\Z}{\mathbb{Z}}

\newcommand*{\canonclass}[0]{\ensuremath K_B}

\newcommand*{\locus}[1]{\ensuremath\{#1\}}

\newcommand*{\divclass}[1]{\ensuremath [#1]}
\newcommand*{\gaugedivclass}[0]{b}

\newcommand{\conjecture}{Automatic Enhancement Conjecture}

\newcommand*{\vev}[0]{VEV}
\newcommand*{\mpb}[0]{\hat{b}} 
\newcommand*{\trpla}[0]{\ensuremath \eta_a}
\newcommand*{\trplb}[0]{\ensuremath \eta_b}
\newcommand*{\ta}[0]{t_{(3)}}
\newcommand*{\tb}[0]{t_{(2)}}
\newcommand*{\tc}[0]{t_{(1)}}
\newcommand*{\td}[0]{t_{(0)}}

\newcommand*{\chargedsinglet}[1]{\ensuremath \bm{\left(\pm#1\right)}} 
\newcommand*{\twochargedsinglet}[2]{\ensuremath \bm{\left(#1,#2\right)}}

\newcommand{\nikhil}[1]{\footnote{\textcolor{orange}{\textbf{NR:\ #1}}}}
\newcommand{\wati}[1]{\footnote{\textcolor{blue}{\textbf{WT:\ #1}}}}
\newcommand{\andrew}[1]{\footnote{\textcolor{red}{\textbf{AT:\ #1}}}}

\newcommand{\clean}{
\renewcommand{\nikhil}[1]{}
\renewcommand{\wati}[1]{}
\renewcommand{\andrew}[1]{}
}

\clean

\title{\boldmath Automatic Enhancement\\ in 6D Supergravity and
F-theory Models}

\author[1]{Nikhil Raghuram,}
\author[2]{Washington Taylor,}
\author[3]{and Andrew P. Turner}

\affiliation[1]{Department of Physics\\Robeson Hall, 0435\\Virginia Tech\\850 West Campus Drive\\Blacksburg, VA 24061, USA}
\affiliation[2]{Center for Theoretical Physics\\Department of Physics\\Massachusetts Institute of Technology\\77 Massachusetts Avenue\\Cambridge, MA 02139, USA}
\affiliation[3]{Department of Physics and Astronomy\\University of Pennsylvania\\ Philadelphia, PA 19104, USA}

\emailAdd{nikhilr at vt.edu}
\emailAdd{wati at mit.edu}
\emailAdd{turnerap at sas.upenn.edu}

\preprint{MIT-CTP-5248}

\abstract{
We observe that in many F-theory models, tuning a specific
gauge group $G$ and matter content $M$ under certain circumstances
leads to an automatic enhancement to a larger gauge group $G'
\supset G$ and matter content $M' \supset M$.  We propose that this
is true for any theory $G, M$ whenever there exists a containing theory $G',
M'$ that cannot be Higgsed down to $G, M$.  We give a number of examples
including non-Higgsable gauge factors, nonabelian gauge factors, abelian gauge
factors, and exotic matter.
In each of these cases, tuning an F-theory model with the desired
features produces either an enhancement or an inconsistency, often
when the associated anomaly coefficient becomes too large.
This principle applies to a variety of models in
the apparent 6D supergravity swampland, including some of the simplest cases
with $\U(1)$ and $\SU(N)$ gauge groups and generic matter, as well as infinite
families of $\U(1)$ models with higher charges presented in the prior
literature, potentially ruling out all these apparent swampland theories.
}

\begin{document}
\maketitle
\flushbottom

\section{Introduction}
Six-dimensional supergravity is a promising arena in which to
understand the relationship between possible low-energy theories of
quantum gravity and UV completions in the context of string theory. In
ten dimensions, it has been shown that all supersymmetric theories of
gravity coupled to gauge fields in flat space-time that satisfy
anomaly cancellation and all other known quantum consistency
conditions are realized, at the level of massless spectra, through
some version of string theory \cite{Adams:2010zy, KimShiuVafaBranes}.
There has been recent progress towards showing that a similar ``string
universality'' principle holds
for supergravity theories in eight (as well as nine and seven)
space-time dimensions \cite{Montero:2020icj, Cvetic:2020kuw}.  In six
dimensions, the space of theories becomes significantly more rich: six
is the largest space-time dimension in which matter fields can arise in
representations other than the adjoint for a supersymmetric quantum
theory of gauge fields and gravity.  The space of 6D supergravity
theories is, however, still reasonably manageable. For the most
part, the theory consists of a set of distinct continuous branches
connected by various transitions into a single large meta-moduli space.
The massless spectra in these different branches are strongly
constrained by gravitational, gauge-gravitational and pure gauge
anomaly conditions.

F-theory \cite{VafaF-theory, MorrisonVafaI, MorrisonVafaII} provides a
powerful nonperturbative tool for exploring a large part of the global
meta-moduli space of 6D supergravity theories in terms of the
geometric moduli spaces of elliptic Calabi--Yau threefolds, which are
linked together in a large connected network \cite{GrassiMinimal,
KumarMorrisonTaylorGlobalAspects, Taylor:2011wt, TaylorHodge} by transitions that
can change the number of tensor multiplets \cite{Seiberg:1996vs,
MorrisonVafaII}, as well as standard Higgs transitions and more
exotic ``matter transitions'' \cite{AndersonGrayRaghuramTaylorMiT}.
There is a close connection between the elliptic Calabi--Yau geometry of
F-theory and the structure of the string charge lattice and anomaly
coefficients in the corresponding low-energy 6D supergravity theory,
making it possible in many branches of the theory to clearly identify
the geometric ingredients needed for the UV completion through
F-theory of the given branch of the low-energy 6D supergravity moduli
space (see, e.g.,\cite{KumarMorrisonTaylorGlobalAspects}).  It was
conjectured in \cite{KumarTaylorStringUni6D} that string
universality should hold for the massless spectra of 6D supergravity
theories, so that any massless spectrum associated with a
supersymmetric theory of gauge fields and gravity not realized in
string or F-theory should be provably inconsistent.  At this point in
time, however, despite ongoing progress both in understanding the
range of possible F-theory constructions and in identifying new
quantum consistency conditions for 6D supergravity theories, there are
still many theories, including some infinite classes of theories, in
the apparent {\it swampland} \cite{VafaSwamp}\footnote{Note: the term
``swampland'' is used slightly differently by different authors.
Here, we refer to the ``apparent swampland'' as those theories that
satisfy all known quantum consistency conditions for theories
coupled to quantum gravity but have no known string realization; a theory
can be removed from the apparent swampland if a string realization
is found, or if a new quantum consistency constraint is identified
that rules out the theory.
A recent review of work on swampland related problems, mostly focusing
on questions separate from the 6D supergravity issues addressed here,
can be found in \cite{PaltiSwampland}.
} of 6D low-energy
supergravity theories that obey anomaly cancellation as well as all
other known quantum consistency conditions and yet have no known
F-theory realization.

In this paper we find a unifying principle that
underlies a range of the theories in the current apparent 6D
supergravity swampland, and we formulate a precise conjecture for a class
of theories that seem to be impossible to realize in F-theory despite
satisfying the known 6D quantum consistency conditions.
This paper was motivated by the observation that certain
apparently consistent $(\SU(3) \times \SU(2) \times \U(1))/\Z_6$ models
cannot be realized in F-theory by a universal Weierstrass realization
identified earlier by the authors \cite{RaghuramTaylorTurnerSM}.
Instead, the rank of the gauge group is forced to increase when the
anomaly coefficients associated with the gauge factors are
sufficiently large.  This kind of enhancement had also been seen in
simple $\SU(2)$ models \cite{MorrisonParkU1,
MorrisonTaylorCompleteness}.  Further exploration has revealed that
indeed many models that seem consistent but resist an F-theory
realization fit with this same general pattern.
In this paper, we generalize these observations in the formulation of a conjecture,
and we elaborate on range of different situations where this kind of automatic
enhancement occurs.
If the general conjecture we present here and its stronger forms for
specific gauge groups hold for general 6D supergravity theories it
would rule out many theories in the apparent 6D supergravity swampland.

The structure of this paper is as follows: In \cref{sec:conjecture},
we formulate the central conjecture, which essentially states that
given any
6D supergravity theory that has a gauge group and matter content
contained within those of another consistent theory that can be
realized in F-theory, but where the former theory cannot be reached by
a Higgs deformation of the latter theory, the former theory does
not arise in F-theory.  In \cref{sec:swamp}, we give a brief
review of some of the main types of theories that appear currently to
be in the 6D supergravity swampland.  In  the following two sections, we
go through a number of examples where we can
demonstrate explicitly that the
conjecture holds for standard F-theory constructions.
\cref{sec:nonabelian} addresses a variety of examples with nonabelian
gauge theories, including non-Higgsable clusters and theories with
gauge group $\SU(N)$, and \cref{sec:abelian} describes examples with
an abelian gauge group, including in particular some infinite families of
apparently consistent theories with gauge group $\U(1)$ and
arbitrarily large charges that were previously known.
In \cref{sec:local}, we give a slightly stronger characterization of the
conjecture in terms of positivity/effectiveness for a variety of
the specific
simple cases studied, and also give a sufficient local condition in some cases
for a theory to either be enhanced or broken in the F-theory
realization.
In \cref{sec:exotic-discrete}, we consider some more complicated
cases, focusing in particular on theories
with exotic nonabelian matter representations
and those with discrete gauge groups. The discussion includes situations in which the
application of the conjecture formulated here is less clear but there
are interesting related questions.
In \cref{sec:lower-dimensions}, we briefly consider how the ideas
of this paper may be extrapolated to lower-dimensional supergravity
theories such as $\cN = 1$ and $\cN = 2$ 4D supergravity
theories, and in \cref{sec:further} we make some concluding remarks.

The work in this paper is complemented by a separate but related work
also nearing completion \cite{MorrisonTaylorCompleteness} that
addresses questions of completeness of the charge spectra and
massless charge spectra in 6D supergravity theories and their F-theory
realizations.

\section{Statement of the conjecture}
\label{sec:conjecture}

The basic proposal of this paper is the following conjecture for 6D
F-theory and supergravity theories:

\newtheorem*{AEC}{Automatic Enhancement Conjecture}
\begin{AEC}
    Given any anomaly-consistent 6D supergravity theory with gauge group $G$ and
    massless matter hypermultiplets in a set of representations $M$, such that
    there exists another anomaly-consistent theory with gauge group $G'\supset G$
    and matter representations $M'\supseteq M$ that is realized in F-theory but
    cannot be broken through a supersymmetry-preserving Higgsing process to the
    theory with gauge group $G$, the theory with gauge group $G$ cannot be
    realized in F-theory. Equivalently, trying to tune the gauge group and matter
    content $G, M$ in an F-theory Weierstrass model leads to an automatic
    enhancement of the singularity types to a theory with gauge group and matter
    content $G' \supset G, M' \supseteq M$.
\end{AEC}

\noindent Some comments on this conjecture:
\begin{itemize}
    \item Note that the theory with gauge group and matter content
    $G', M'$ must contain the theory with $G, M$ in the strong sense that
    all other features of the theory must be the same.  In particular, the
    string charge lattice that contains the anomaly coefficients
    (associated with $H_{1,1} (B,\Z)$ of the compactification base $B$ in
    an F-theory description; see, e.g., \cite{KumarMorrisonTaylorGlobalAspects}) and the positivity cone
    on this lattice (associated with the cone of effective divisors in
    F-theory) must be the same, and the anomaly coefficients of the gauge
    factors in $G$ must match their counterparts in $G'$.

    \item We have stated this conjecture in the context of F-theory.  Much of the bulk of this paper consists of a variety of examples where this conjecture holds for standard tunings in the F-theory context.  While in some cases, such as the situation of non-Higgsable clusters (\cref{sec:non-Higgsable}), the conjecture can be explicitly proven, in other situations the evidence is only partial even in the F-theory context.  Nonetheless, it is tempting to speculate that this conjecture holds not only in F-theory but for general 6D supergravity theories.  If so, this would eliminate many of the known examples of 6D supergravity theories in the {\it swampland} of theories that appear consistent by all known criteria but which are not realized in F-theory.

    \item This conjecture complements the {\it Massless Charge
    Sufficiency Conjecture} made in \cite{MorrisonTaylorCompleteness},
    which asserts that the massless charge lattice of a 6D supergravity
    theory generates the full charge lattice of the theory.  As discussed
    further in \cref{sec:su2-quotient}, in some cases such as the enhancement of a
    theory with gauge algebra $\asu(2)$ with a large anomaly coefficient,
    the massless charge sufficiency conjecture is closely related to the
    automatic enhancement conjecture and can be similarly interpreted as
    leading to the addition of extra elements in the root lattice of the
    gauge group $G$.

    \item On a practical level, enhancements typically occur because the choice of divisor classes necessary to obtain the correct matter spectrum causes the elliptic fibration to take some special structure. A parameter could be forced to be zero because its divisor class is ineffective, or multiple parameters may share common factors because their classes only admit reducible sections. And if a parameter's divisor class is trivial, the elliptic fibration may exhibit enhanced features such as the presence of additional rational sections. Since these effects occur in the geometry, one may naively wonder if the $G$,$M$ model could be obtained from the $G^\prime$, $M^\prime$ model by a T-brane deformation \cite{CecottiCordovaHeckmanVafaTBranes,AndersonHeckmanVafaTBranes}. However, examples of the \conjecture\ exhibit a field-theoretic obstruction to Higgsing $G^\prime$ to $G$. T-brane deformations should respect these obstructions, implying that automatic enhancement cannot be evaded in this fashion.

    \item The conjecture as stated above corresponds to the complete
    6D supergravity theory and its F-theory realizations, and is stated in
    a very general way.  Most of the
    examples we describe here also admit a
    more specific and stronger formulation in a way that depends on the
    particular group/algebra in question.  In
    particular, in many cases the enhancement mechanism is
    required when the anomaly
    coefficients of the relevant gauge factors theory violate a particular
    positivity condition, associated with an effectiveness condition for
    the corresponding F-theory curves.  When this condition is violated
    either an enhancement must occur or there cannot exist a compatible
    F-theory model.  In many cases these positivity constraints lead to
    sufficient conditions that can be formulated completely locally in the
    F-theory geometry.  We illustrate  these positivity and local aspects in many
    of the examples described in the following sections, and summarize
    these versions of the conjecture in \cref{sec:local}.
\end{itemize}

\section{Review of 6D supergravity, F-theory realizations, and the swampland}
\label{sec:swamp}

In this section we briefly review some aspects of the current apparent
``swampland'' of 6D supergravity theories that appear consistent from
all known quantum consistency conditions but that have no known
F-theory realization.

\subsection{6D supergravity theories and their realizations in F-theory}

A 6D $\cN = 1$ supergravity theory is characterized by a massless
spectrum that contains a fixed number of tensor multiplets $T$, a
gauge group $G$, and a set of hypermultiplet matter fields $M$ living
in certain representations of the gauge group $G$.  The gauge group
can have nonabelian and abelian factors, as well as discrete
components.  The conditions of gravitational, gauge-gravitational, and
pure gauge anomaly cancellation \cite{SagnottiGS,
GreenSchwarzWest6DAnom} impose fairly strong constraints on the set
of possible gauge groups and massless matter spectra.  When the number
of tensor multiplets satisfies $T <9$, it has been proven that there
are a finite number of distinct possible massless spectra $G, M$ for
theories with purely nonabelian gauge groups $G$
\cite{KumarMorrisonTaylorGlobalAspects}.  On the other hand, even
when there are no tensor multiplets $(T = 0)$, there exist infinite
families of theories with a simple $\U(1)$ gauge group and different
combinations of charged matter fields \cite{ParkTaylorAbelian,
TaylorTurnerSwampland} that satisfy anomaly cancellation and other
known consistency conditions.

A wide range of 6D $\cN = 1$ supergravity theories can be
realized using F-theory \cite{VafaF-theory, MorrisonVafaI,
MorrisonVafaII}.  A six-dimensional F-theory model is defined by an
elliptic Calabi--Yau threefold $X$, and can be thought of as a
nonperturbative type IIB compactification on the (real
four-dimensional) complex K\"ahler surface $B$ that acts as the base
of the elliptic fibration $X$.  Introductions to the relevant aspects
of F-theory compactifications can be found in \cite{Taylor:2011wt,
WeigandTASI}.  The set of 6D F-theory models forms a connected
meta-moduli space; the allowed bases are connected through
nonperturbative transitions that change the number of tensor
multiplets \cite{Seiberg:1996vs, MorrisonVafaII}, and over any given
base $B$ the set of valid F-theory models forms a continuous connected
moduli space parameterized by the coefficients $f, g$ in the
Weierstrass model
\begin{equation}
 y^2 = x^3 + fx + g \,.
\end{equation}
Here $f, g$ can be thought of as functions on the base $B$, or more
precisely as sections of line bundles $\cO (-4\canonclass), \cO (-6
\canonclass)$, where $\canonclass$ is the canonical class of the base.  Codimension-one loci where the elliptic fiber becomes singular lead to nonabelian
factors of the gauge group $G$, while abelian and discrete factors
have a more subtle global characterization.  Codimension-two loci
where the elliptic fiber becomes singular in general characterize
massless matter fields in the 6D supergravity theory.  There is thus a
close correspondence between the geometry of the F-theory
compactification and the structure of the low-energy 6D supergravity
theory.

One particularly important connection between low-energy physics and
F-theory is that there is a simple geometric interpretation of the
anomaly coefficients $a, b_i, \tilde{b}_\alpha$ associated with terms
in the 6D supergravity action of the form $B\wedge R\wedge R, B\wedge
F_i\wedge F_i, B\wedge F_\alpha\wedge F_\alpha$ respectively, with
$F_i, F_\alpha$ denoting nonabelian and abelian gauge field strength
factors. In particular, $a$ corresponds to the canonical class $\canonclass$,
and $b_i$ corresponds to the locus $C_i$ in $B$ where the nonabelian
gauge factor $G_i$ is supported.  Note that for $k$ abelian factors
there are anomaly coefficients $\tilde{b}_{\alpha \beta}$ associated
with terms $B \wedge F_\alpha \wedge F_\beta$; we use the notation
$\tilde{b}_\alpha = \tilde{b}_{\alpha \alpha}$.  These are special
cases of the more general identification between the string charge
lattice $\Lambda$ of the 6D supergravity theory, in which $a, b_i,
\tilde{b}_\alpha$ are elements, and the lattice $H_{1, 1}(B,\Z)$ given
by the set of algebraic curves in the base $B$, including $-\canonclass$ and
$C_i$, with the natural intersection form on homology.  Because of
this isomorphism, we will frequently move freely back and forth
between $a, b_i \in\Lambda$ and $K_B, C_i \in H_{1, 1}(B,\Z)$ in
various parts of the discussion.  A useful feature of
this correspondence is that the arithmetic genus of a (possibly
singular) curve $C_i$ supporting a nonabelian gauge factor $G_i$ is
encoded in the anomaly coefficients of the low-energy theory through
\cite{KumarParkTaylorT0, KleversEtAlExotic}
\begin{equation}
\label{eq:genus}
g_i =\frac{1}{2} (a \cdot b_i + b_i \cdot b_i) + 1\,.
\end{equation}

To organize our brief review of some classes of models in the 6D
supergravity swampland, we will find it helpful to distinguish
``generic'' from exotic matter fields for a given gauge group.  A
precise definition of which matter fields are generic for a 6D
supergravity theory with  a fixed gauge group $G$ (and for fixed
anomaly coefficients, which are assumed to be not too large)
can be given by identifying the generic matter fields as those that
arise on the branch of moduli space of greatest dimension for fixed
$G$ and associated anomaly coefficients \cite{TaylorTurnerGeneric}.
This definition of generic turns out to match with the simplest types
of codimension-two singularities that give matter in the F-theory
picture and also ties in naturally to the structure of the anomaly
constraints: at least for simple theories, the number of generic
matter representations is equal to the number of anomaly constraints,
so in these cases there is a unique generic matter content for fixed
$G$ once the anomaly coefficients are fixed.

Another  aspect of F-theory constructions that is relevant for the analysis of
this paper is that for a given
gauge algebra $\aG$, the moduli space of models with matter fields in
the generic
representations  is captured in many cases by a
``universal''\footnote{In \cite{RaghuramTaylorTurnerSM},  these universal
Weierstrass model constructions were referred to as ``generic'';
here we change terminology to ``universal'' to avoid confusion with
other uses of the term generic.}  Weierstrass model, with a general
algebraic form parameterized by the classes of the curves supporting
the gauge factors.  In \cite{RaghuramTaylorTurnerSM} we developed a
systematic Jacobian-based methodology for checking that a given family of Weierstrass
model constructions is universal by checking that the number of
independent free complex parameters in the Weierstrass model matches the dimensionality of the branch of the moduli
space determined by the number of uncharged scalars.  Such universal
Weierstrass models include, for example, the well-known Tate tunings
of most nonabelian gauge factors \cite{BershadskyEtAlSingularities},
the Morrison--Park $\U(1)$ Weierstrass models \cite{MorrisonParkU1},
and the  universal $(\SU(3) \times \SU(2) \times \U(1)) / \Z_6$ models
developed in \cite{RaghuramTaylorTurnerSM}.  Note that universal
Weierstrass models are also known for some classes of models
containing specific combinations of representations that include
exotic matter.

\subsection{The 6D supergravity swampland}

While for many classes of theories there is a fairly close match
between the set of possible F-theory constructions and the 6D $\cN = 1$ supergravity theories that are allowed from anomaly
cancellation and other known quantum consistency conditions, there are
also many theories that appear to be in the swampland.  In
\cite{KumarMorrisonTaylorGlobalAspects, ParkTaylorAbelian, TaylorTurnerSwampland}, for
example, some infinite families of apparent swampland models were
identified.  In general, both supergravity and F-theory models are
understood best for theories with no tensor multiplets ($T = 0$), and
with nonabelian gauge groups and generic matter representations.  The
structure of the theories becomes more complicated as the number of
tensor multiplets increases, particularly at $T \ge 9$, when abelian
and discrete gauge factors are included, and when exotic (non-generic)
matter representations are considered.  Over the last decade or so,
there has been gradual progress in better understanding both F-theory
constructions of these more complicated classes of models and also in
identifying further quantum consistency constraints that must be
satisfied by any consistent $\cN = 1$ 6D supergravity theory; as
these theories are better understood, further classes of apparent
swampland models have also been identified.  We briefly review some of
this work here, first summarizing some additional constraints on 6D
supergravity and F-theory constructions, and then focusing on some of
the simplest and largest classes of apparent swampland models, for
which the \conjecture\ is relevant.

\subsubsection{Constraints on 6D supergravity theories}
\label{sec:constraints}

In addition to the constraints of local and global anomaly
cancellation, a consistent 6D $\cN = 1$ supergravity theory must
satisfy a variety of further quantum consistency conditions.  We
briefly summarize some of these further constraints here.

One simple constraint is that the gauge kinetic term proportional to
$F^2$ must have a negative coefficient; otherwise the theory has a
perturbative instability.  This leads through
supersymmetry to the condition that the anomaly coefficients $b_i,
\tilde{b}_\alpha$ must lie in the positive cone of the 6D string charge
lattice $\Lambda$.  This was used, for example, in
\cite{KumarTaylorBound} to rule out an infinite family
of models that had appeared to lie in the swampland \cite{SchwarzAnomaly}.

Another constraint, proven in \cite{SeibergTaylorLattices}, is that the charge
lattice $\Lambda$ must be unimodular (self-dual)
under the Dirac pairing between
strings; this follows automatically from F-theory, where the lattice
$H_{1, 1}(B,\Z)$ is always unimodular, but  this
condition must also hold more generally in 6D supergravity, ruling out other classes of
potential swampland models.

Some further constraints on the anomaly coefficients $a, b_i,
\tilde{b}_\alpha$ were identified in \cite{MonnierMooreParkQuantization,
MonnierMooreGS} using global and local anomaly cancellation
conditions (a subset of these constraints were also found in
\cite{Raghuram34, TaylorTurnerSwampland}).  These constraints are
stronger if one makes the additional assumption that the 6D theory can
be defined on any spin manifold with an arbitrary gauge bundle.  For
example, under these assumptions the anomaly coefficient $a$ must be a
{\it characteristic vector} in the subspace of $\Lambda$ spanned by
the nonabelian anomaly coefficients $b_i$, meaning that $a \cdot b_i +
b_i \cdot b_i \in 2\Z$. This is true for all
F-theory compactifications (since, e.g., the genus \labelcref{eq:genus} is
always an integer), but can rule out some theories that might
appear to be in the swampland.

There are a number of further constraints that are known to hold in
F-theory but have not  been proven as consistency conditions for general
quantum 6D supergravity theories.
Theories satisfying the known 6D supergravity quantum consistency
conditions but violating these F-theory constraints appear to be in
the swampland, pending a proof that these conditions are more
generally necessary for 6D supergravity consistency.  The primary goal
of this paper is to present evidence that the automatic enhancement
conjecture should be a similar F-theory constraint.  We summarize
briefly here some other constraints in this class.
Note that beyond standard F-theory models there are also other
possible constructions such as the ``frozen'' phase of F-theory
\cite{Bhardwaj:2018jgp}, which can realize some  models that otherwise
may appear to be in the swampland.  As far as we know the models we
study here do not have any such realization through any formulation of
string theory including the standard class of F-theory constructions.

One constraint that follows directly from F-theory is the \emph{Kodaira
constraint} \cite{KumarMorrisonTaylorGlobalAspects}, which states
that $-12a-\sum_{i}\nu_ib_i$ must lie in the positivity cone of
$\Lambda$, which corresponds to the cone of effective divisors in
$H_{1, 1} (B,\Z)$. Here $\nu_i$ corresponds to the number of 7-branes
needed to form the nonabelian factor $G_i$ of the gauge group,
e.g., for $\SU(N), \nu = N$.  Various 6D supergravity theories that
look otherwise consistent violate this condition and hence cannot be
realized in F-theory; we mention some explicit examples later in this
section.

Another constraint that is widely expected to hold in all consistent
quantum theories of gravity in dimensions $D >3$ is the condition of
charge completeness \cite{Banks:2010zn}, which states that there exist
excitations of the theory with all possible charges in the weight
lattice of the symmetry group $G$. This was proven recently for all
quantum gravity theories with a holographic dual
\cite{Harlow:2018tng}.  This condition is proven for all 6D F-theory
models in \cite{MorrisonTaylorCompleteness} using Poincar\'{e}
duality, although there is no general proof for flat space 6D
supergravity theories.

A further class of constraints can be identified if one assumes that
the charge completeness hypothesis holds in the sense that there are
dynamical string excitations for every charge in the string lattice
$\Lambda$.  Under this assumption, there must be a consistent
anomaly-free theory on the worldvolume of such strings, which leads to
additional constraints on the set of allowed spectra for 6D
supergravity theories.  This method was used in \cite{Banks:1997zs},
for example, to show that the rank of the gauge group of 10D type I
string theory must match with the known value of $\aso(32)$.  In
\cite{KimShiuVafaBranes}, this approach was used to show that
consistency of the theory on the string worldvolume places additional
constraints on the spectra of 6D supergravity theories; among other
things, these constraints would rule out several infinite families of
apparent swampland models with $T > 9$ that were found in
\cite{KumarMorrisonTaylorGlobalAspects}.  In
\cite{LeeWeigandSwampland}, this approach was used to place bounds on
the number of abelian gauge factors that can arise in a 6D
supergravity theory, improving on previous bounds found from anomaly
conditions in
\cite{ParkTaylorAbelian}.

The
focus in this paper is on 6D supergravity models that obey all of the
above-mentioned constraints but still have no F-theory realization.
To organize the discussion we summarize some of the models that are
still in the apparent swampland, including some that violate the
conditions above that are known to hold in F-theory vacua but are
otherwise apparently consistent 6D supergravity theories.

\subsubsection{Apparent swampland for $T = 0$, nonabelian $G$, and
generic matter}

We begin with the simplest set of cases, with no tensors, a
nonabelian gauge group, and generic matter.  Because of the absence of
tensor fields, the anomaly coefficients $a, b_i$ are integers.  We
focus first on theories with gauge group $\SU(N)$ and then consider
other gauge groups.  For $G =\SU(N)$, generic matter fields are the
fundamental and adjoint representation, as well as the two-index
antisymmetric representation when $N \ge 4$
\cite{TaylorTurnerGeneric}.

With gauge group $\SU(2)$, there are anomaly-allowed theories for
anomaly coefficients $1 \le b \le
12$ \cite{TaylorTurnerGeneric}.  Naively, it seems that these
theories can all be constructed directly in F-theory by directly tuning the
appropriate gauge factor in the Weierstrass model.  When $b = 9, 10,
11, 12$, however, complications arise.  In the case $b = 12$, as
discussed further in \cite{MorrisonTaylorCompleteness}, this
construction gives a theory with $\Z_2$ Mordell--Weil torsion, so the
gauge group is actually $\SO(3)$; in this case, the massless charge
sufficiency conjecture stated in that paper  asserts that the pure
$\SU(2), b = 12$ theory cannot be realized in F-theory.
For $b = 10, 11$, it was observed in \cite{MorrisonTaylorCompleteness}
that the naive tuning of this model gives an extra $\SU(2)$ factor and
Mordell--Weil torsion so $G = (\SU(2) \times \SU(2))/\Z_2$.  These
cases, as well as the case $b = 9$, are in the apparent swampland and
satisfy the conditions of the \conjecture, and are described in more
detail in \cref{sec:su2su2} and \cref{sec:su2u1}.

For gauge group $\SU(N)$, there are anomaly-allowed theories for $1
\le b \le 24/N$.  Again, naively these can all be constructed using
a Tate tuning \cite{BershadskyEtAlSingularities}.  But again, in some
cases the resulting model does not have the desired $\SU(N)$ gauge
group.  It was pointed out in \cite{MorrisonTaylorMaS,
JohnsonTaylorEnhanced} that for $b = 1$ this construction does not
give theories with gauge groups $\SU(21)$ and $\SU(23)$ and generic
matter, so these are in the apparent swampland.  These cases, as well
as the other cases with $N \ge 18$ are examples of the \conjecture,
as we describe in more detail in \cref{sec:t0-general-n}.  Note that
the case of $\SU(24)$, which lacks fundamental matter and has an
actual gauge group $\SU(24)/\Z_2$ is an example where Mordell--Weil
torsion is forced, in accord with the Massless Charge Sufficiency
Conjecture \cite{MorrisonTaylorCompleteness}.


Similar apparent swampland theories arise for $\asp(N)$ and $\aso(N)$; we discuss some cases in \cref{sec:spsoalgebras}.

\subsubsection{Apparent swampland with abelian $G$}

When we include abelian $\U(1)$ gauge factors, the story becomes more
complicated.   In F-theory, $\U(1)$ factors come from global features
of the geometry, which makes them more subtle to understand.
For theories with a gauge group $\U(1)$ and generic $(q = 1, 2)$
matter, there are apparent swampland models closely related to the
simplest $\SU(2)$ models discussed above; we describe in
\cref{sec:abelian-generic} how these models also satisfy the \conjecture.
Even for theories with a single $\U(1)$ as the gauge
group and $T = 0$, when larger charges are considered
 there are infinite families of charge configurations
that satisfy anomaly cancellation conditions
\cite{ParkTaylorAbelian,TaylorTurnerSwampland}, even though there can only be a finite
number of distinct spectra compatible with F-theory constructions
\cite{KumarMorrisonTaylorGlobalAspects}.  In \cref{sec:abelian-infinite}, we
show that at least the simplest of these families satisfies the
conditions of the \conjecture.

Combining nonabelian and abelian factors, one can also find infinite
families of anomaly-free theories with gauge algebras of the form
$\asu(N) \oplus \au(1)$, for example. (Some infinite families of this
type with $T = 1$ were constructed in \cite{ParkTaylorAbelian}, and
similar constructions are possible with $T = 0$.)  In
\cref{sec:abelian}, we show that the previously identified infinite
classes of theories of this kind satisfy the conditions of the
stronger version of the \conjecture\ that would render these theories
inconsistent, although we also identify some further infinite families
for which the application of the conjecture is less
transparent. We also show in
this section that the \conjecture\ applies to some theories with
gauge group $(\SU(3) \times \SU(2) \times \U(1)) / \Z_6$ when
constructed with the universal Weierstrass form identified in
\cite{RaghuramTaylorTurnerSM}.

Going beyond single $\U(1)$ factors, at this point our understanding
of F-theory models with multiple $\U(1)$ factors and/or discrete gauge
group factors is still sufficiently limited that it is difficult to
clearly delineate the swampland.  This is in large part because
abelian $\U(1)$ and discrete factors come from global aspects of the
F-theory geometry, unlike nonabelian gauge factors that come from
local structure.  We do not attempt to give any systematic analysis
here of theories with multiple $\U(1)$ factors or discrete gauge
factors, though we make some general comments about theories with
discrete gauge symmetries in \cref{sec:discrete}.

\subsubsection{Apparent swampland with $T > 0$}
\label{sec:tg0}

When the number of tensor multiplets is nonzero, $T>0$, the structure
of the theory becomes more complicated, and questions about the
swampland have additional complications.  The anomaly coefficients $a,
b_i, \tilde{b}_\alpha$ become vectors as  $\Lambda$ becomes a signature
$(1, T)$ lattice, corresponding to $H_{1, 1}(B,\Z)$.  There is a
positivity cone for strings in $\Lambda$, which in the F-theory
description corresponds to effective divisors in $B$ but which is not
as well understood in the 6D supergravity context.  A variety of
apparent swampland models may arise corresponding to lattices
$\Lambda$ with positivity cones that do not match any F-theory
geometry.  The condition identified in
\cite{MonnierMooreParkQuantization} that $a$ must be a characteristic
vector, at least for the nonabelian anomaly coefficients, puts some
concrete constraint on the allowed combinations of $\Lambda, a$, but
there are still many theories at larger $T$ that appear consistent and
yet have no F-theory realization.  We focus here on some of these that
are most relevant for this story.

The most basic new feature that can arise at $T > 0$ is the appearance
of a vector $x \in\Lambda$ that lies in the positivity cone and
satisfies $x \cdot x <0$.  When an F-theory construction involves a
base $B$ that contains an effective curve $C$ with $C \cdot C <- 2$,
the negative normal bundle on $C$ forces a Kodaira singularity in the
elliptic fibers over that curve that gives a gauge group of at least
$\SU(3)$ in the corresponding 6D supergravity theory
\cite{MorrisonTaylorClusters}.  From the point
of view of supergravity, there is no obvious inconsistency with an
element $x$ satisfying $x \cdot x <-2$ lying in the positivity cone
without a corresponding gauge factor in the theory.  Thus, such models
lie in the apparent swampland.  In \cref{sec:non-Higgsable}, we show
that these models are a simple example of the \conjecture.  Note that
under the assumption of charge completeness for the string charge
lattice, one could
alternatively argue
that anomaly cancellation on the worldvolume of a string with string
charge $x$ can only be satisfied in the presence of the appropriate
non-Higgsable gauge factor $G\supset\SU(3)$, using analysis along the
lines of \cite{Haghighat:2014vxa, Kim:2016foj, Shimizu:2016lbw, KimShiuVafaBranes}.

Already with a single tensor multiplet $(T = 1)$, the  range of
possible theories expands significantly and there are more theories
that have been identified in the apparent swampland.
In particular, in \cite{ParkTaylorAbelian} there are infinite families
of apparent swampland models identified with $T = 1$ and gauge algebras
$\au(1)$ and $\asu(13) \oplus \au(1)$.  Both of these families  can be
ruled out by the \conjecture, as we discuss in \cref{sec:abelian}.
The $\U(1)$ family is closely related to the infinite family of
$\U(1)$ theories at $T = 0$ found in \cite{TaylorTurnerSwampland}, but
the $\asu(13) \oplus\au(1)$ family is somewhat more subtle.

When $T  \ge 9$, the classification of theories gets even more
complicated, due in particular to the fact that $a \cdot a = 9-T$
ceases to be positive (this condition follows from gravitational
anomaly cancellation).
An illustration of some of the complications that arise at $T \ge 9$
is given in  \cref{app:adjoint}.
For $T \ge 9$, there is no proof known that even constrains the
number of distinct spectra for nonabelian gauge groups and matter to
be finite, though as mentioned earlier some of the infinite families of
apparent swampland models identified in
\cite{KumarMorrisonTaylorGlobalAspects} violate the conditions found
in \cite{KimShiuVafaBranes} that hold when string charge completeness
is assumed.

\subsubsection{Apparent swampland with exotic  matter for nonabelian $G$}

In addition to the generic matter representations for a nonabelian gauge group
$G$ there are also non-generic matter representations possible that
can still satisfy anomaly cancellation.  Three-index antisymmetric
representations of $\SU(N)$ for $N = 6, 7, 8$ are fairly well
understood \cite{BershadskyEtAlSingularities, MorrisonTaylorMaS,
GrassiMorrisonAnomalies, AndersonGrayRaghuramTaylorMiT} and can
arise through a Tate-type tuning of the Weierstrass model
\cite{HuangTaylorComparing}.  Related exotic multi-charged matter
representations include, e.g., the $(\bm{6}, \bm{2})$ of $\SU(4) \times\SU(2)$
and the trifundamental of $\SU(2)$. More exotic
matter representations can be characterized as having a non-vanishing
``genus'' contribution \labelcref{eq:genus}, corresponding to the arithmetic genus of a
singularity in the divisor supporting the gauge factor in an F-theory
realization.  In \cite{KleversEtAlExotic}, it was argued that only a
fairly limited set of exotic matter representations for nonabelian groups
can be realized in
F-theory without having a codimension two locus where $f, g$ vanish to
degrees (4, 6), which generally signifies a superconformal sector
\cite{Seiberg:1996qx, Heckman:2013pva, DelZotto:2014hpa}.  These
limited higher-genus exotic  matter representations that can be
realized through F-theory models on a singular gauge divisor basically
contain only the two-index symmetric representation of $\SU(N)$, the
three-index symmetric representation of $\SU(2)$,and analogous
representations of $\Sp(N)$.  It has been suggested that some other
exotic representations such as the $(\bm{56},  \bm{2})$ of $\gE_7 \times \SU(2)$,
which naively involves a (4, 6) locus, may be possible without a
superconformal sector \cite{Cvetic:2018xaq} through T-brane magic.
Though this is not well understood in F-theory, it is believed that there are
heterotic models containing this matter representation without
superconformal sectors.

With this general classification of exotic matter, there are two kinds
of apparent swampland theories.  First, there are anomaly-consistent
theories containing the exotic matter representations that can arise
from F-theory constructions on singular divisors, but in
configurations that do not seem to arise from F-theory.  Some examples
of such apparent swampland theories were described in
\cite{KleversEtAlExotic}; we show in \cref{sec:exotic-F-theory} that
some such examples satisfy the conditions of the \conjecture.  Second,
are any theories that contain matter in representations of nonabelian
gauge factors not believed to arise in consistent F-theory models.  We
make some brief comments on such models in
\cref{sec:exotic-non-F-theory}.

\section{Nonabelian Examples}
\label{sec:nonabelian}

\subsection{Non-Higgsable Clusters}
\label{sec:non-Higgsable}

One of the simplest situations in which automatic
enhancement must occur in F-theory is when $G$ is trivial and there is
a primitive positive charge $q$ in the string charge lattice with
Dirac pairing $q \cdot q <-2$ with itself and $q \cdot a =- q \cdot q
  -2$ with the $B R R$ anomaly coefficient $a$.  In any F-theory
construction where the complex surface base $B$ on which the theory
is compactified contains a curve $C$ of self-intersection $-3$ or
below, everywhere in the F-theory moduli space
of elliptically fibered Calabi--Yau threefolds parameterized by
Weierstrass models over $B$
there is a
singularity in the elliptic fibration that forces at least a gauge
algebra $\asu(3)$ with an anomaly coefficient $b= C$, where the
string charge lattice containing the anomaly coefficient is
identified with $H_{1,1} (B,\Z)$ in any 6D F-theory model
\cite{MorrisonTaylorClusters}.  Thus, in such a theory with a
primitive positive charge $b$ where $b \cdot b = -3$, $a \cdot b =
1$ there is an automatic enhancement $G = 1 \to G' \supseteq \SU(3)$.
 the self-intersection $b \cdot b$ decreases, the rank of the forced
gauge group increases, with, e.g., $b \cdot b = -4 \Rightarrow G'
\supseteq \SO(8)$, \textellipsis, $b \cdot b < -8 \Rightarrow G' = \gE_8$, and
with $b \cdot b \le -13$ not allowed in any F-theory model.
In all these situations, the gauge group $G'$ is ``non-Higgsable'' in the
sense that there is no local Higgsing deformation that can break the
gauge group while preserving supersymmetry.
In particular, in these
cases there is no Higgsing from $G' \to G$, as expected in the
\conjecture.

Note that in this case the \conjecture\ can be stated purely locally:
we expect that any charge $b$ in the string charge lattice $\Lambda$
that satisfies $b \cdot b < -2$ and satisfies $a \cdot b = -2-b \cdot
b$ (corresponding to $g = 0$ from \cref{eq:genus}) must support a
gauge factor $G' \supseteq \SU(3)$, with correspondingly larger
factors for more negative $b \cdot b$.  Furthermore, for similar
reasons there cannot be any a curve in the positivity cone with $g >
0$ and $b \cdot b < 0$ in any F-theory construction
  \cite{MorrisonTaylorClusters}.

In this class of automatic enhancements there is a complete proof that
the enhancement must occur in the context of F-theory.  In other
cases, our understanding of the F-theory geometry is less complete and
while we can give many examples of standard tunings that lead to
automatic enhancement, the proof that this must always occur even in
the F-theory context is less complete.  Whether the automatic
enhancement associated with non-Higgsable clusters can be proven
purely in the 6D supergravity context through other approaches is also
discussed briefly in \cref{sec:tg0}.

\subsection{Automatic enhancement of $\asu(2)$ gauge algebra}
\label{sec:su2-enhancement}

Another very simple set of examples of the kind of automatic
enhancement implied by the \conjecture\ arises in the case of 6D
F-theory models with gauge group $\SU(2)$, generic matter (fundamental
and adjoint representations), and no tensor multiplets (associated
with the base $\bP^2$), where the anomaly coefficient for the
gauge group is taken to be $b=10,11$.  This was one of the first
examples of the automatic enhancement mechanism observed, and was
encountered in \cite{MorrisonParkU1, MorrisonTaylorCompleteness}. We
describe how the automatic enhancement occurs in F-theory for this
kind of theory and then describe some further generalizations thereof.

\subsubsection{$\asu(2) \to (\SU(2) \times\SU(2))/\Z_2$
automatic enhancement}
\label{sec:su2su2}

A universal form for the Weierstrass model giving a gauge algebra $\asu(2)$
and the generic fundamental and adjoint matter representations
can be constructed  either from the
Tate form \cite{BershadskyEtAlSingularities, KatzEtAlTate}, or by direct
analysis of the general Weierstrass model with the appropriate Kodaira
singularity type \cite{MorrisonTaylorMaS}.
\footnote{We assume here that the $\asu(2)$ algebra is realized
through a UFD-type construction. Non-UFD constructions
typically give exotic matter representations
\cite{KleversEtAlExotic}. Although in principle it
may be possible that in
some special cases non-UFD models can be realized with only generic
matter types, the analysis of \cref{app:u1u1test} suggests that this
will not occur in practice.}
Such a universal  Weierstrass model with
gauge algebra $\asu(2)$ realized on a curve $\sigma = 0$
takes the form
\begin{equation}
\label{eq:su2}
\begin{aligned}
    f & = -\frac{1}{48} \phi^2 + \sigma f_1\,, \\
    g &=\frac{1}{864} \phi^3 -\frac{1}{12} \phi \sigma f_1 + \sigma^2 ({g}_2)\,.
\end{aligned}
\end{equation}
The classes of $f, g, \sigma$ in the F-theory base are $[f] = -4 K_B,
[g] = -6 K_B, [\sigma] = C$, where $K_B$ is the canonical class of the
base.  In the 6D supergravity theory, $K_B$ and $C$ correspond to the
anomaly coefficients $a, b$ of the $B R^2, B F^2$ terms respectively;
we freely go back and forth between the geometric (curve in $H_{1, 1}
(B,\Z)$) and field theory (anomaly coefficient in the 6D supergravity
string charge lattice) notations $K_B \leftrightarrow a, C
\leftrightarrow b$ throughout the discussion in this and subsequent
sections.  The discriminant locus associated with the Weierstrass
model \labelcref{eq:su2} is
\begin{equation}
\Delta = \frac{1}{16}  \phi^2 (\phi g_2-f_1^2)  \sigma^2 +\cO (\sigma^3)\,.
\end{equation}

The class of the coefficient $g_2$ is $[g_2] = -6 a -2 b$.  When $b$
is sufficiently large, this class becomes ineffective and $g_2 = 0$.
In such a situation, the discriminant becomes
$\Delta = \sigma^2 f_1^2 (-\phi^2 + 64f_1 \sigma)/16$.  This
discriminant is divisible by $f_1^2$, and so
there is an automatic enhancement of {\it at least}
an extra $\asu(2)$ algebra on the
vanishing locus of $f_1$, where $[f_1] = -4a-b$.  As shown in
\cite{MorrisonTaylorCompleteness}, the resulting Weierstrass model
also has $\Z_2$ torsion, so the gauge group becomes $G' =(\SU(2)
\times\SU(2))/\Z_2$.
The fact that the enhanced gauge group arises automatically from the
Weierstrass tuning
(which is the universal and minimal (UFD) tuning needed for the $\asu(2)$ algebra) implies that there is no Higgsing from the $G'$
theory to the anomaly-free theory with algebra $\asu(2)$.
This corresponds to the absence of adjoint matter fields under the
extra $\asu(2)$ algebra factor on $f_1 = 0$
(for further explanation/discussion on this point, see \cref{app:adjoint}).

\paragraph{Simple examples}

The simplest cases in which this enhancement occurs is when
the base is $\bP^2$ and the anomaly coefficient is $b = 10$ or $b =
11$.   In these cases an additional $\asu(2)$  factor arises on $f_1$
with $[f_1] = 2, 1$ respectively.
Since degree one and two curves on $\bP^2$ are rational curves of
genus 0, the additional factor carries no adjoint matter.
Furthermore, the only fundamental matter fields for the additional
$\asu(2)$ are also charged under the original $\asu(2)$,\footnote{This
can be seen from the form of the discriminant locus; the only loci
over $f_1= 0$ where the vanishing order of $\Delta$ increases are
those where either $\sigma = 0$, corresponding to matter jointly
charged under both $\asu(2)$ factors, and those where $\phi = 0$,
corresponding to an enhancement to a Kodaira type III singularity that
carries no matter.} so the
enhanced $G' = (\SU(2) \times\SU(2))/\Z_2$ cannot be Higgsed to the
anomaly-free $\SU(2)$ theory on $b = 10, 11$, matching the conditions
of the \conjecture.
Note that there are Higgsings of the $G'$ theory down to a theory with
gauge algebra $\asu(2)$; in particular, Higgsing on the bifundamental
fields can break these $G'$ theories  down to an $\SO(3)$ on $b = 12$
with only adjoint matter essentially by deforming $f_1 \sigma
\to \sigma'$ where $\sigma'$ is a generic degree 12
curve in $\bP^2$.   Such a Higgsing, however, cannot
reproduce the original $b = 10, 11$ theory that acquired the automatic enhancement.

This kind of $\asu(2)$ enhancement can also occur on many other
F-theory bases, not just on $\bP^2$.  On the Hirzebruch surface
$\F_m$, for example,  if $\sigma = 0$ is $n$ times the curve $\tilde{S}$ of positive
self-intersection $+m$, then in a basis where $S$ is the curve of
self-intersection $- m$ and $F$ is the fiber curve with $F \cdot F =
0, F \cdot S = 1$ (such that $\tilde{S}= S + m F$), the class of $g_2$
is
\begin{equation}
    [g_2] =-6 K_B -2 C = 6 (2 S + (2+ m) F) - 2n (S + m F)
    = (12 - 2n) S + (12+6 m - 2n m) F \,.
\end{equation}
This becomes ineffective when $n > 3+6/m$.  So, for example, on the
Hirzebruch surface $\F_3$, tuning an $\asu(2)$ gauge algebra on $6
\tilde{S}$ would force an additional $\asu(2)$ to arise on the curve
$f_1 = 0$, where $[f_1] = -4 K_B - C = 2S+ 2F$.  Part of this
enhancement leads to an enhancement of the $\SU(3)$ gauge factor on
the $-3$ curve $S$, since $[f_1]$ contains $2S$ as a component.
This can be seen (using the Zariski decomposition  as in
\cite{MorrisonTaylorClusters})
by noting that $[\sigma] \cdot S= 0, [\phi] \cdot
S= -2, [f_1] \cdot S= -4$, so the degrees of vanishing of $f, g,
\Delta$ on $S$ are $2, 3, 6$ and the gauge factor on $S$ is enhanced
to $\SO(7)$.  The remaining part of
$[f_1]$ is $[f_1] -2S= 2F$; this is a reducible curve, giving two
$\asu(2)$ factors each on a copy of $F$, so the total gauge group of
the enhanced theory is
\begin{equation}
    G' = (\SU(2) \times \SU(2) \times \SU(2) \times \SO(7))/\Z_2 \,.
\end{equation}
The matter content includes 6 bifundamentals between the first
$\SU(2)$ factor and each of the other $\SU(2)$ factors, and a
half-hypermultiplet in the $(\bm{2}, \bm{8})$ between each of
the enhanced $\SU(2)$ factors and the $\SO(7)$ factor (analogous to
the matter appearing in the $-2, -3, -2$ non-Higgsable cluster
\cite{MorrisonTaylorClusters}).
It is clear that there is no Higgsing of this model possible that
would leave the initial $\SU(2)$ factor unbroken.
Many other similar models can be realized on other bases with
non-Higgsable gauge factors.

\paragraph{Positivity condition and local formulation}

The condition under which the automatic $\asu(2)$ enhancement
mechanism enhancement must occur is that the divisor $[g_2] = - 6 a -2
b$ is ineffective, corresponding to the condition in the 6D
supergravity theory that $- 6 a -2 b$ is not in the (closure of the)
positivity cone in $\Lambda$.
Thus in the case of $\asu(2)$ a stronger formulation of the conjecture
asserts that an $\asu(2)$ factor with anomaly coefficient $b$,
where $-6a-2b$ is not in the positivity cone,
is not
possible unless there is also at least another $\asu(2)$ factor  with
anomaly coefficient $b' = -4a-b$.
The ``local'' information about the
$\asu(2)$ algebra is encoded in the quantities $a \cdot b, b \cdot b$,
or equivalently $g, b \cdot b$, corresponding in the F-theory picture
to the genus (through \cref{eq:genus}) and self-intersection of the
curve $C$ supporting the desired $\asu(2)$ factor.  When $b \cdot b
>0$, for example, it is then a sufficient condition for $[g_2]$ to be
outside the positivity cone that $- 3 a \cdot b <b \cdot b$.  As an
example of the local version of $\asu(2)$ enhancement, we can thus
conjecture that whenever a theory contains an $\asu(2)$ algebra with
an anomaly coefficient $b$ satisfying $b \cdot b >0, b \cdot b >-3 a
\cdot b$ there must also be a nonabelian gauge factor of rank at least
one with an anomaly coefficient $b'= [f_1] = -4 a - b$, or the theory
cannot be consistent at least with any F-theory construction.  Note
that when this $b'$ vanishes there is no additional factor, but there
is an additional section as discussed in \cref{sec:su2-quotient}.
When $b'$ is not in the (closure of the) positivity cone, then the
Weierstrass model develops codimension one (4, 6) loci and is
inconsistent.  This local formulation clearly matches with the
examples above; for example, when $T = 0, -a = 3$, the extra $\asu(2)$
factor arises when $b = 10, 11$.  Finally, note that on the F-theory
side it is possible that $f_1$ is reducible, so that the additional
gauge factor may be present with a pair (or more) of anomaly
coefficients $b_1, \ldots, b_k$ with $b' = b_1 + \cdots + b_k$.

\paragraph{Enhancements that break the theory}
\label{sec:break}

Note that there are also cases where a similar forced enhancement in
the Weierstrass model can lead to codimension one (4, 6) loci, which
break the F-theory construction; these cases give swampland models
that are not addressed by the global version of the automatic
enhancement conjecture as stated in \cref{sec:conjecture}, but which
are addressed by the more specific positivity condition and local
version stated above.  As a simple example of this
consider an $\asu(2)$ factor on the curve $C =4 \tilde{S}$ in the
Hirzebruch surface base $\F_{12}$.  In this case, $[g_2] = 4S-12F$ is
ineffective and $g_2$ vanishes.  We furthermore have $[f_1] = 4S+ 8F$,
so $f_1$ vanishes to order $4$ on $S$.  Since $\phi$ vanishes to order
2 on $S$, there is a codimension one (4, 6) singularity on $S$ so no
F-theory model is possible.  One way of understanding this is that the
automatic enhancement on any curve intersecting or containing a $-12$
curve would naively lead to matter charged under the $\gE_8$, which is
not possible.  This model fits with the local version of the $\asu(2)$
automatic enhancement conjecture as stated above, since $b \cdot b =
192, -a \cdot b = 56$, so $-3a \cdot b = 168 < b \cdot b$, so the
local version of the conjecture rules out the theory with gauge
group $\gE_8 \times \SU(2)$ with anomaly coefficients $S, 4 \tilde{S}$.

\subsubsection{$\asu(2) \to (\SU(2) \times \U(1))/\Z_2$
automatic enhancement}
\label{sec:su2u1}

An interesting further class of related examples occurs with the above
$\asu(2)$ tuning when $[g_2] = 0$, so  $g_2$ is a constant.  In this
case we get an automatic enhancement to a theory with an additional
$\U(1)$ factor.  The story is similar to the preceding enhancement
that gives the additional $\SU(2)$ factor.  One way to understand the
resulting $(\SU(2) \times \U(1))/\Z_2$ theory is to consider the theory
with gauge group $(\SU(2) \times\SU(2))/\Z_2$ that would arise with
$[g_2] = 0$, so $g_2$ is a constant.  In this situation, $[f_1] = -4a-b = -a$, so $[f_1]$ is a
genus one curve; as in the cases above, all fundamental matter in the
extra $\asu(2)$ factor on this curve comes from bifundamentals with
the original $\asu(2)$.  Thus, we can Higgs this theory on the adjoint
of the second $\SU(2)$, giving a theory with gauge group
\begin{equation}
\label{eq:g21}
G' =(\SU(2) \times \U(1))/\Z_2 \,,
\end{equation}
where all $\U(1)$-charged matter is in the $\bm{2}_{1/2}$
representation.  Thus, the $G'$ theory cannot be Higgsed further to a
pure $\SU(2)$ theory and this represents another class of automatic
enhancements satisfying the conjecture.

This argument is somewhat implicit.  A more explicit analysis
(see \cref{app:ratsections})
shows
that when $g_2$ is a constant, the Weierstrass model takes the form of
a Morrison--Park model \cite{MorrisonParkU1} with the additional
$\SU(2)$ factor on $\sigma = 0$.
As for the enhancements described above, this story continues to hold
in the presence of non-Higgsable gauge factors on more complicated
bases, though often the presence of the $\U(1)$ factor simply
gives $\U(1)$ charges to the non-Higgsable gauge factors and does not
enhance them further.
The simplest example of this enhancement arises on the base $\bP^2$
when an $\asu(2)$ factor is tuned on a curve of degree $b = 9$.

Note that this this class of enhancement conditions is similar
to that described in \cref{sec:su2su2}, but occurs in conditions
where $-3a = b$ holds identically.  This is not really a local
condition, as $a \leftrightarrow K_B$ involves global information about the geometry
in F-theory, as would be expected when abelian $\U(1)$ factors are
involved.

\subsubsection{$\asu(2) \to \SO(3)$ and massless charge sufficiency}
\label{sec:su2-quotient}

Note that a similar mechanism to the \conjecture\ operates
when we tune an $\asu(2)$ algebra on a curve $\sigma$ in the
class $-4a$.  In such a situation, $[f_1] = 0$ and there is no new gauge factor. But as shown in
\cite{MorrisonTaylorCompleteness},
in F-theory there is automatically a new $\Z_2$ element in
Mordell--Weil torsion, which gives the gauge group the global structure
$\SO(3) = \SU(2)/\Z_2$.  While this is not an automatic enhancement,
this is a case of the Massless Charge Sufficiency condition described
in \cite{MorrisonTaylorCompleteness}.  For example, on the base
$\bP^2$, tuning an $\asu(2)$ algebra on curves of
degrees $9, 10, 11$ leads in F-theory to an automatic enhancement
through the mechanism described above, while tuning an $\asu(2)$
algebra on a curve of degree 12 leads in F-theory to an $\SO(3)$
theory.  Thus, if both the \conjecture\ and the
Massless Charge Sufficiency Conjecture are correct in 6D
supergravity, then there would be no swampland for 6D $\cN = 1$
theories with $T = 0$, gauge group $\SU(2)$, and generic (fundamental
and adjoint)
matter.

There is a sense in which the \conjecture\ and the Massless Charge
Sufficiency Conjecture are special cases of a common phenomenon.
 The appearance of the new element of Mordell--Weil torsion in
the $\asu(2) \to \SO(3)$ situation can in some sense be thought of as adding a new element
to the root lattice of $G$, within the linear span of $G$.  In this
sense, the principal difference between the \conjecture\ and the
Massless Charge Sufficiency Conjecture is whether the additional root
added  increases the dimensionality of the gauge algebra or lies in
the same vector space.  This interpretation is very clear in the
F-theory context, where these roots correspond to homology classes in
the compactification space.
It would be interesting to investigate whether a similar
interpretation can be given  directly in 6D supergravity that would
unify these principles.

The local version of this class of conditions is similar
to that described in \cref{sec:su2su2}, but occurs in conditions
where $-4a = b$ holds identically, so $b' = [f_1] = 0$.
(Note that if $-4a-b = (-a) + (-3a-b)$ = 0 then $-3a-b$ cannot lie in
the positivity cone since $-a$ does for all F-theory models.)

\subsection{$T=0$ models with large $\asu(N)$ gauge algebras at $b = 1$}
\label{sec:t0-general-n}

Similar enhancements occur in models with $\asu(N)$ gauge algebras for
$N>2$. Let us consider models with $T=0$ tensor multiplets, an
$\asu(N)$ gauge group with a $b=1$ anomaly coefficient, and the
following spectrum of  charged hypermultiplets:\footnote{For $N=3$, the hypermultiplet spectrum would consist of 24 fundamental hypermultiplets, but the $N=3$ case is not of significant interest here.}
\begin{equation}
\label{eq:sunspectrum}
3\times \bm{\frac{N(N-1)}{2}} + (24-N)\times\bm{N}\,.
\end{equation}
Such massless spectra are consistent with the anomaly cancellation
conditions for $N \le 24$; for larger $N$, we would require a negative
number of fundamental hypermultiplets.  Note that these
representations (fundamental and two-index antisymmetric), along with
the adjoint, are the generic matter representations for
$\asu(n)$.

There are also consistent $T=0$ models with $\asu(N)\oplus\au(1)$
gauge algebras and generic matter
where the anomaly coefficient for the $\asu(N)$ factor
is $b=1$, at least for $N<19$. For even $N$, the spectrum of charged
hypermultiplets is
\begin{equation}
\label{eq:sunu1spectrumeven}
3\times \left(\bm{\frac{N(N-1)}{2}}\right)_{0} + \left(12-\frac{N}{2}\right)\times \bm{N}_{\frac{1}{2}} +  \left(12-\frac{N}{2}\right)\times \bm{N}_{-\frac{1}{2}} + \left(9-\frac{N}{2}\right)\left(12-\frac{N}{2}\right)\times\bm{1}_{1}\,,
\end{equation}
while for odd $N$, the spectrum of charged hypermultiplets is
\begin{equation}
\label{eq:sunu1spectrumodd}
3\times \left(\bm{\frac{N(N-1)}{2}}\right)_{-\frac{1}{N}} + \frac{19-N}{2}\times \bm{N}_{-\frac{N+1}{2N}} +  \frac{29-N}{2}\times \bm{N}_{\frac{N-1}{2N}} + \frac{1}{4}\left(19-N\right)\left(23-N\right)\times\bm{1}_{1}\,.
\end{equation}
Although these spectra naively agree with \cref{eq:sunspectrum} if we ignore the $\au(1)$ charges, the additional $\au(1)$ gauge algebra is clearly an important feature. In general, the gauge algebra can be Higgsed down to $\asu(N)$ by giving \vev{}'s to two hypermultiplets of $\bm{1}_{1}$ matter. But if $N$ is either 18 or 19, there are no $\bm{1}_{1}$ hypermultiplets, and the Higgsing is not permitted. The \conjecture\ therefore implies that the $\asu(18)$ or $\asu(19)$ models above should not occur in F-theory, as the gauge algebra should enhance to $\asu(18)\oplus\au(1)$ or $\asu(19)\oplus\au(1)$.

This enhancement can be seen explicitly if we attempt to construct
these models. Because there are no tensor multiplets, the
corresponding F-theory models would have $\bP^2$ as the base of
the elliptic fibration. And to obtain an $\asu(N)$ gauge algebra with
a $b=1$ anomaly coefficient, the models should have split $I_{N}$
singularities tuned along a line in this $\bP^2$. There are
known UFD Weierstrass tunings that realize split $I_N$ singularities
along a divisor $\locus{\sigma=0}$ \cite{MorrisonTaylorMaS,
KatzEtAlTate}. The Weierstrass model takes the form
\begin{equation}
\label{eq:weierstrasssuneven}
y^2 = x^3 + \left(-\frac{1}{3}\Phi^2 + f_{k}\sigma^{k}\right)x z^4 + \left(\frac{2}{27}\Phi^3 - \frac{1}{3}\Phi f_{k}\sigma^{k} + g_{N}\sigma^{N}\right)z^6
\end{equation}
for $N = 2 k$ and
\begin{equation}
\label{eq:weierstrasssunodd}
\begin{aligned}
    y^2 &= x^3 + \left(-\frac{1}{3}\Phi^2 + \frac{1}{2}\phi_0 \psi_k \sigma^k + f_{k+1}\sigma^{k+1}\right)xz^4 \\
    &\qquad\qquad + \left(\frac{2}{27}\Phi^3 -\frac{1}{6}\Phi\left(\phi_0\psi_k+2f_{k+1}\sigma\right)\sigma^k + \frac{1}{4}\psi_k^2\sigma^{2k} + g_N \sigma^N\right)z^6
\end{aligned}
\end{equation}
for $N=2k+1$, where\footnote{In \cite{MorrisonTaylorMaS}, $\Phi$ is defined as $\frac{1}{4}\phi_0^2 + \phi_1 \sigma + \ldots + \phi_{k-1}\sigma^{k-1}$. However, one can remove the $\phi_{i>1}$ parameters by the redefinition $\phi_{1}\to \phi_{1} - (\phi_{2}\sigma + \ldots + \phi_{k-1}\sigma^{k-2})$.}
\begin{equation}
\label{eq:Phituning}
\Phi = \frac{1}{4}\phi_0^2 + \phi_1 \sigma\,.
\end{equation}
These models support hypermultiplets in the antisymmetric and
fundamental representations.\footnote{If $\locus{\sigma=0}$ is a curve
of genus greater than 0, there may also be adjoint hypermultiplets,
although this possibility is not important here.} Thus, the natural
strategy for constructing our desired $\asu(N)$ models is to apply
these tunings in an elliptic fibration over $\bP^2$ with
$\divclass{\sigma}=H$.
The other parameters then have the following
divisor classes:
\begin{equation}
\divclass{\phi_{0}} = 3H\, \quad \divclass{\phi_{1}} = 5H\, \quad \divclass{\psi_k} = (9-k)H\, \quad \divclass{f_i} = (12-i)H\, \quad \divclass{g_j} = (18-j)H\,.
\end{equation}

For $N\le 17$ and $N=24$, these tunings give us the desired $\asu(N)$ models without any enhancements of the gauge algebra.\footnote{The $N=24$ model, however, is an example of the Massless Charge Sufficiency Conjecture, as an additional torsional section appears.} But when $N=18$, $\divclass{g_N}$ is trivial, and $g_N$ is automatically a perfect square. If we replace $g_N$ with $\gamma^2$, the elliptic fibration described by \cref{eq:weierstrasssuneven} admits a rational section
\begin{equation}
\label{eq:sungenseceven}
[\secx:\secy:\secz] = \left[\frac{1}{3}\Phi: \gamma \sigma^{9}: 1\right]\,.
\end{equation}
The gauge algebra is therefore $\asu(18)\oplus\au(1)$, and one can
confirm with the Katz--Vafa method \cite{KatzVafa} and the proposals
in \cite{RaghuramTurnerOrders} that the charged hypermultiplet
spectrum agrees with \cref{eq:sunu1spectrumeven}. When $N=19$,
$\divclass{g_N}$ is ineffective, and $g_N$ is essentially zero. After
setting $N$ to 19 and $g_N$ to 0, the elliptic fibration described by
\cref{eq:weierstrasssunodd} admits the rational section
\begin{equation}
\left[\secx:\secy:\secz\right] = \left[\frac{1}{3}\Phi:\frac{1}{2}\psi_{9}\sigma^{9}:1\right]\,,
\end{equation}
implying that the gauge algebra is enhanced to
$\asu(19)\oplus\au(1)$. And by the Katz--Vafa method \cite{KatzVafa}
and the proposals in \cite{RaghuramTurnerOrders}, one can again verify
that the charged hypermultiplet spectrum agrees with
\cref{eq:sunu1spectrumodd}. Clearly, we see the enhancements expected
from the \conjecture\ in these two models.

Enhancements also occur for $N$ larger than 19, as previously observed in \cite{MorrisonTaylorMaS, JohnsonTaylorEnhanced}.\footnote{Because it is somewhat more difficult to find extra rational sections, these previous works did not point out the additional $\au(1)$ algebras for $N=18$ and $19$.} The enhancements are summarized in \cref{tab:sunenhancements}. If we use \cref{eq:weierstrasssuneven} to construct the $\asu(20)$ and $\asu(22)$ models, the divisor class $\divclass{g_N}$ becomes ineffective, and the gauge algebras enhance to $\asu(20)\oplus\asu(2)$ or $\asu(22)\oplus\asu(2)$ respectively. (The generating section in \cref{eq:sungenseceven} becomes torsional \cite{BaumeEtAlTorsion}, and the gauge group is $(\SU(N)\times\SU(2))/\Z_{2}$.) While the extra $\asu(2)$ algebra in both cases could be Higgsed by giving \vev{}s to two $(\bm{1},\bm{2})$ hypermultiplets, the $\asu(20)\oplus\asu(2)$ and $\asu(22)\oplus\asu(2)$ spectra in \cref{tab:sunenhancements} do not have this matter. Thus, the appearance of this extra $\asu(2)$ algebra agrees with the \conjecture.

Meanwhile, if we use  \cref{eq:weierstrasssunodd} to construct the
$\asu(21)$ and $\asu(23)$ models, the divisor class
$\divclass{\psi_{k}}$ becomes ineffective, and the gauge algebras
enhance to $\asu(22)\oplus\asu(2)$ and $\asu(24)$,
respectively. Neither of these enhanced models has the requisite
matter to Higgs the gauge algebra to $\asu(21)$ or $\asu(23)$. While
one can give \vev{}s to $(\bm{22},\bm{2})$ matter to break
$\asu(22)\oplus\asu(2)$ to $\asu(21)$ \cite{LiGroupTheory}, one would
need at least two hypermultiplets of $(\bm{22},\bm{2})$ matter
to satisfy the D-term constraints
\cite{DanielssonStjernberg}.
And the $\asu(24)$ model, which supports only antisymmetric
hypermultiplets, lacks the fundamental matter necessary to Higgs the
gauge algebra down to $\asu(23)$. Therefore, the observed enhancements
when attempting to construct the $\asu(21)$ and $\asu(23)$ models
agree with the expectations of the \conjecture.

The local version of this class of enhancements states that for any
element $b \in\Lambda$ satisfying $b \cdot b = 1, g = 0$ ($-a \cdot b =
3$), an isolated gauge algebra of $\asu(18)$ or more is automatically
enhanced according to \cref{tab:sunenhancements}.

\begin{table}
    \centering

    \begin{tabular}{*{3}{c}}\toprule
    Expected Gauge Algebra & Enhanced Gauge Algebra & Matter Spectrum \\ \midrule
    $\asu(18)$ & $\asu(18)\oplus\au(1)$ & $3\times\bm{153}_{0} + 3\times\bm{18}_{1/2}+ 3\times\bm{18}_{-1/2}$\\
    $\asu(19)$ & $\asu(19)\oplus\au(1)$ & $3\times\bm{171}_{-1/19} + 5\times\bm{19}_{9/19}$\\
    $\asu(20)$ & $\asu(20)\oplus\asu(2)$ & $3\times(\bm{190},\bm{1})+2\times(\bm{20},\bm{2})$\\
    $\asu(21)$ & $\asu(22)\oplus\asu(2)$ & $3\times(\bm{231},\bm{1})+(\bm{22},\bm{2})$\\
    $\asu(22)$ & $\asu(22)\oplus\asu(2)$ & $3\times(\bm{231},\bm{1})+(\bm{22},\bm{2})$\\
    $\asu(23)$ & $\asu(24)$ & $3\times\bm{276}$\\
    $\asu(24)$ & $\asu(24)$ & $3\times\bm{276}$\\ \bottomrule
    \end{tabular}

    \caption{Enhanced gauge algebras observed when attempting to tune various expected gauge algebras along a divisor of class $H$ on $\bP^2$, along with the observed matter spectrum.}
    \label{tab:sunenhancements}
\end{table}

\subsection{Other $\SU(N)$ examples}

In the last two subsections we have focused on enhancements of the
algebra $\asu(2)$ with general anomaly coefficients and enhancements
of the algebra $\asu(n)$ at large $n$ on anomaly coefficients $b \cdot
b=1$,
where $g= 0$.  We briefly discuss some more general $\asu(n)$ examples

\subsubsection{Enhancements of $\asu(3)$ and beyond for general
  $b$}

A similar enhancement to that described in \cref{sec:su2su2} occurs
for $\asu(3)$.  When the coefficients in classes $[f_2] = -4a-2b$ and
$[g_3]= -6a-3b$ in a $\sigma$ expansion of $f, g$ for the general
$\SU(3)$ form \cite{MorrisonTaylorMaS, MorrisonTaylorCompleteness} are
not effective, the discriminant acquires an additional cubic factor
$\psi_1^3$ where $[\psi_1] = -3a-b$.  This gives an enhancement
$\asu(3) \to \asu(3) \oplus \asu(3)$.  Just as there is $\Z_2$ torsion
in the $\asu(2)\to\asu(2)\oplus\asu(2)$ situations of
\cref{sec:su2su2}, there is automatically $\Z_3$ Mordell--Weil torsion in these
cases \cite{MorrisonTaylorCompleteness}, so the global gauge group
becomes $G' = (\SU(3) \times \SU(3))/\Z_3$.  Again, the second $\SU(3)$
factor has no adjoints and all fundamentals/antifundamentals are
jointly charged, so the second $\SU(3)$ factor cannot be Higgsed
without breaking the original $\SU(3)$.  For $T = 0$, this enhancement
occurs for the values $b =7, 8$.  At $b =  6$, $f_2, g_3$  become constants
and the algebra is enhanced by $\au(1) \oplus \au(1)$, by the same logic that led to
\cref{eq:g21}.  The local version of this analysis is that when $b
\cdot b > 0, b \cdot b > -2a \cdot b$ we get an extra $\asu(3)$ factor
on $-3a-b$.

We now consider $\asu(4)$.  In general what is needed for automatic
enhancement to an extra $\SU(2)$ factor with generic (fundamental,
adjoint) matter that cannot be Higgsed down is that the extra factor
must have an anomaly coefficient $b'$ giving $g' = 0$ (for the absence
of adjoints) and the automatically enhanced group must provide all
fundamental $\SU(2)$ matter.  The number of fundamental matter fields
in this situation is $b' \cdot (-8a-2b')$ from anomaly cancellation.
For an $\asu(4)$ with anomaly coefficient $b$ to experience
enhancement to $ \asu(4)\oplus \asu(2)$ we must then have $4b \cdot b'
= b' \cdot (-8a-2b')$.  This can certainly occur, most simply when $2b
= -4a-b'$.  For example, this occurs when $T = 0$ where $b = 5, b' =
2$.  In general, this will occur when $-3a-2b$ is not positive but
$-4a-2b$ is positive; it is not hard to confirm in the universal
Weierstrass model for an $\asu(4)$ with generic matter that this
enhancement occurs under these conditions, where $f_3 = g_4 = 0$.

We leave further exploration of automatic enhancement  of this type
for gauge factors with general anomaly coefficients to further work.

\subsubsection{Enhancements of $\asu(N)$ for $g = 0,$ small $b \cdot b$}
\label{sec:small-bb}

We now focus on more general enhancements of $\asu(N)$ for larger $N$,
restricting to the cases where $g = 0$. For $\asu(N)$ models with $g =
0$, $b \cdot b = 2$, the largest value of $N$ allowed by the anomaly
cancellation conditions is $N=16$. Automatic enhancement should occur
for at least $N\ge 13$ because of the requirement that $(-6a-Nb)\cdot
b$ must be non-negative for
$\divclass{g_N}= -6a-Nb$ to be effective. But automatic enhancement
can also occur for some $N$ less than 13 depending on the base of the
F-theory model in question, since $-6a-N b$ can be outside the
positivity cone even if $(-6a-N b) \cdot b \ge 0$.  For instance, let
us restrict attention to $T=1$ models, which are constructed using the
Hirzebruch surfaces $\F_m$ as bases. To ensure that the
divisor supporting the $\asu(N)$ algebra is irreducible and has
self-intersection $+2$, $m$ should be $0$ or $2$.  The automatic
enhancements, which are the same for both of these cases, are
summarized in \cref{tab:selfint2enhancements}.

The pattern of enhancements may be different for $b\cdot b=2$ divisors
on other bases. For instance, consider a base $B$ found by blowing up
$\F_2$ at the intersection of the $-2$ curve and a $0$ curve,
and imagine tuning an $\asu(N)$ algebra on  a $+2$ curve $\locus{\sigma=0}$. There are no automatic
enhancements for $N\le9$ on $B$, apart from a relatively
uninteresting $\asu(3)$ NHC on the $-3$ curve. For $N=10$,
$\divclass{g_{10}} = -6\canonclass-10\divclass{\sigma}$ is effective
and nontrivial, but the only holomorphic sections of
$\divclass{g_{10}}$ are perfect squares. The gauge algebra should
therefore admit an extra $\au(1)$ factor. In fact, the full gauge
algebra is $\asu(10)\oplus\asu(3)\oplus\asu(2)\oplus\au(1)$, and the
charged hypermultiplet spectrum is
\begin{equation}
    4\times(\bm{45},\bm{1},\bm{1})_{0} + 1\times(\bm{10},\bm{1},\bm{2})_0+ 5\times (\bm{10},\bm{1},\bm{1})_{\frac{1}{2}}+5\times(\bm{10},\bm{1},\bm{1})_{-\frac{1}{2}}\,.
\end{equation}
This spectrum satisfies the conditions of the \conjecture, as one
cannot Higgs away the $\asu(2)$ or $\au(1)$ factors without disturbing
the $\asu(10)$ algebra. Tuning $\asu(10)$ on the $b\cdot b=2$ curve in
$B$ should therefore lead to an automatic enhancement as observed,
even though a similar $\asu(10)$ enhancement does not occur in the
$T=1$ models above.
Locally, this enhancement can be understood as being forced from the
intersection of the $+ 2$ curve with a curve of self-intersection
$-1$, and can be understood using purely local analysis as in
\cite{JohnsonTaylorEnhanced}.
Likewise, $\asu(11)$ and $\asu(12)$ experience
automatic enhancements when tuned on the $+2$ curve of $B$, yet they
can be obtained without enhancements when $T=1$.

\begin{table}
    \centering

    \begin{tabular}{*{3}{c}} \toprule
        Expected Gauge Algebra & Enhanced Gauge Algebra & Matter Spectrum \\ \midrule
        $\asu(12)$ & $\asu(12)\oplus\au(1)$ & $4\times\bm{66}_{0}+4\times\bm{12}_{\frac{1}{2}}+4\times\bm{12}_{-\frac{1}{2}}$\\
        $\asu(13)$ & $\asu(13)\oplus\au(1)$ & $4\times\bm{78}_{-\frac{1}{13}} + 6\times\bm{13}_{\frac{6}{13}}$\\
        $\asu(14)$ & $\asu(14)\oplus\asu(2)$ & $4\times(\bm{91},\bm{1})+2\times(\bm{14},\bm{2})$\\
        $\asu(15)$ & $\asu(16)$ & $4\times\bm{120}$\\
        $\asu(16)$ & $\asu(16)$ & $4\times\bm{120}$\\ \bottomrule
    \end{tabular}

    \caption{Automatic enhancements for $T=1$ models with $\asu(N)$ gauge algebras tuned on divisors of genus $g=0$ and self-intersection $b\cdot b=2$.}
    \label{tab:selfint2enhancements}
\end{table}

For $g=0$ and $b\cdot b=3$, the largest value of $N$ allowed by the
anomaly cancellation is 13. Based on the condition for $(-6a-Nb)\cdot
b$, automatic enhancements should occur at least for $N\ge11$. Again,
the actual pattern of enhancements depends on the base of the F-theory
model. When $T=1$, one can find irreducible curves with
self-intersection $+3$ on $\F_1$ and $\F_3$. On
$\F_1$, there is no enhancement for $N\le8$, and the enhanced
gauge algebras are $\asu(9)\oplus\asu(2)$ for $N=9$,
$\asu(10)\oplus\asu(2)\oplus\asu(2)$ for $N=10$, and
$\asu(12)\oplus\asp(2)$\footnote{We use conventions where $\asp(2)$ is
supported along non-split $I_4$ loci.} for $N=11$ and $N=12$. On
$\F_{3} $, there is no enhancement for $N\le9$ apart from the
usual $\asu(3)$ non-Higgsable cluster. The enhanced gauge algebras are
$\asu(10)\oplus\au(1)\oplus\asu(3)$ for $N=10$,
$\asu(11)\oplus\au(1)\oplus\asu(3)$ for $N=11$, and
$\asu(12)\oplus\asu(2)\oplus\asu(2)\oplus\aso(7)$ for
$N=12$,\footnote{Since it can be difficult to find rational sections, the enhanced gauge algebras may actually have $\au(1)$ factors beyond those listed here, which we were able to identify easily.} with the last gauge algebra representing that supported
on the $-3$ curve. For both of these bases, attempting to obtain an
$\asu(13)$ gauge algebra using the standard tuning of
\cref{eq:weierstrasssunodd} produces an invalid elliptic fibration
with singular fibers everywhere on the base. While we do not discuss
the matter spectra for brevity, they are all consistent with the
\conjecture.

\subsection{Examples with $\asp(N)$ and $\aso(N)$ algebras}
\label{sec:spsoalgebras}

We can immediately obtain $\asp(m)$ examples of automatic enhancement by Higgsing the $\asu(2m)$ models of \cref{sec:t0-general-n}. If we give a \vev{} to matter in the antisymmetric representation of $\asu(2m)$, we obtain an $\asu(m)$ gauge algebra with the same anomaly coefficient $b$. From an F-theory perspective, the resulting $\asp(m)$ model is given by taking the $\asu(2m)$ construction of \cref{eq:weierstrasssuneven} and ignoring the tuning of $\Phi$ in \cref{eq:Phituning}. The $\asu(2m)$ models of \cref{sec:t0-general-n} have enough antisymmetric hypermultiplets to perform this Higgsing. However, these hypermultiplets are uncharged under any additional $\asu(2)$ or $\au(1)$ factor appearing due to automatic enhancements, and the $\asu(2m) \to \asp(m)$ breaking does not introduce new hypermultiplets that would let us Higgs the extra gauge factors without disturbing the $\asp(m)$ algebra. This fact provides us with $T=0$ examples of automatic enhancement with $\asp(m)$ gauge algebras and $b=1$ anomaly coefficients.

When $m=9$, for instance, the conjecture implies that $\asp(9)$ experiences an automatic enhancement to $\asp(9)\oplus\au(1)$ with the following charged hypermultiplet spectrum:
\begin{equation}
    2\times \bm{152}_{0} + 3\times\bm{18}_{1/2} + 3\times\bm{18}_{-1/2}\,.
\end{equation}
This enhancement can be seen in the $\asp(9)$ Weierstrass model
produced by the procedure described above: the class
$\divclass{g_{18}}$ is trivial, leading to an additional section of
the elliptic fibration. Similarly, $\asp(10)$ and $\asp(11)$
experience automatic enhancements to $\asp(10)\oplus\asu(2)$ and
$\asp(11)\oplus\asu(2)$, respectively. While the $\asp(12)$ model does
not experience automatic enhancement, the Weierstrass model has an
extra torsional section, in agreement with the Massless Charge
Sufficiency Conjecture as noted in
\cite{MorrisonTaylorCompleteness}.

There are also $\aso(N)$ examples with automatic enhancement. For example, consider a $T=0$ model with an $\aso(10)$ gauge algebra and $b=4$ anomaly coefficient. The charged hypermultiplet spectrum for this model is
\begin{equation}
    3\times\bm{45} + 8\times\bm{16}+4\times\bm{10}\,.
\end{equation}
However, there is also a consistent $\aso(10)\oplus\au(1)$ model with anomaly coefficients $b=4$, $\tilde{b}=1$ and a charged hypermultiplet spectrum of the form
\begin{equation}
    3\times\bm{45}_{0} + 8\times\bm{16}_{\frac{1}{4}}+4\times\bm{10}_{\frac{1}{2}}\,.
\end{equation}
The \conjecture\ implies that we should see an $\aso(10) \to \aso(10) \oplus \au(1)$ enhancement. In fact, this can be seen in the $\aso(10)$ Weierstrass tuning
\begin{equation}
    y^2 = x^3 + \sigma^2\left(-\frac{1}{48}\Phi^2 + f_3 \sigma\right)+\sigma^3\left(\frac{1}{864}\Phi^3 + \left(\gamma^2 - \frac{1}{12}f_3\Phi\right)\sigma+g_{5}\sigma^2\right)\,.
\end{equation}
The $\aso(10)$ algebra is tuned along $\locus{\sigma=0}$, so $\divclass{\sigma}$ should be $4H$ to obtain the correct $\aso(10)$ hypermultiplet spectrum. But the divisor class $\divclass{g_5}=-6\canonclass-5\divclass{\sigma}=-2H$ is then ineffective, forcing $g_5$ to be 0. In turn, the elliptic fibration acquires an additional rational section of the form
\begin{equation}
    [\secx : \secy : \secz] = \left[\frac{1}{12}\sigma\Phi : \gamma \sigma^2 : 1\right]\,,
\end{equation}
demonstrating the expected appearance of the $\au(1)$ gauge algebra.

\section{Abelian Examples}
\label{sec:abelian}

The \conjecture\ also suggests that abelian gauge symmetries in certain models
are forced to enhance. In fact, the conjecture is particularly powerful in
these settings, as it can explain why infinite families of apparently
consistent supergravity models with abelian gauge symmetries do not
occur in F-theory. We first describe cases of the \conjecture\ with
$\U(1)$ gauge group and generic ($q = 1, 2$) matter, and then focus on
two types of infinite families that include exotic $\U(1)$ charges:
those with $\U(1)$ gauge groups, and an infinite family with an
$\asu(13)\oplus\au(1)$ gauge algebra.

\subsection{$\U(1)$ models with generic matter}
\label{sec:abelian-generic}

For a gauge group $\U(1)$, the generic matter representations are
those with charges $q = 1, 2$ \cite{TaylorTurnerSwampland}.
Weierstrass models for F-theory constructions of $\U(1)$ theories with
these generic matter charges were constructed by Morrison and Park
\cite{MorrisonParkU1}.  The explicit  form of these Weierstrass models
is given in \cref{eq:Morrison--Park}, and it was  shown that these
models are universal in the sense that they capture all the uncharged
moduli, at least over simple bases, in \cite{WangU1s,
MorrisonTaylorMultisection}.
There is a class of situations where the generic $\U(1)$ model with
certain anomaly coefficients satisfies the conditions of the
\conjecture.  These situations are closely related to the $\asu(2)$
models described in \cref{sec:su2su2}.  Consider a 6D supergravity
theory with gauge algebra $\asu(2) \oplus \asu(2)$, and anomaly
coefficients $b, b'$ satisfying $b + b' = -4 a$.  In such a situation,
the spectrum consists of $b \cdot b'$ bifundamental matter fields and
$g, g'$ adjoint fields for the two $\asu(2)$ factors, where $g, g'$
are given by \cref{eq:genus} using $b, b'$ respectively.  When $g
>0$, the first $\asu(2)$ factor can be broken down to $\au(1)$ by
Higgsing on an adjoint field.  When $g' = 0$, however, the second
$\asu(2)$ factor cannot be broken through Higgsing while preserving
the first $\au(1)$ factor.  This gives a situation with gauge algebra
$\au(1) \oplus \asu(2)$ and matter charged only in the $\bm{1}_2,
\bm{2}_1$ representations.  When such a theory can be constructed in
F-theory, the conditions of the \conjecture\ are satisfied, so we
expect that the theory with only gauge algebra $\au(1)$ and anomaly
coefficient $\tilde{b}= 2 b$ cannot be realized in F-theory.

The simplest example of this is in the case of no tensor multiplets
$(T = 0)$.  There are anomaly-consistent $\U(1)$ models with anomaly
coefficient $\tilde{b}= 6, 8, \ldots, 24$ and spectrum $\tilde{b}(24 -
\tilde{b}) \times\chargedsinglet{1}, \tilde{b}(\tilde{b}- 6)/4 \times\chargedsinglet{2}$. Naively, these come from Higgsing the anomaly-consistent
$\asu(2)$ model with $b = \tilde{b}/2$.  As we have seen, however, in
\cref{sec:su2su2}, the $\asu(2)$ models with $b = 10, 11$ experience
automatic enhancement, and from the above analysis the same is
expected for the $\au(1)$ models with $\tilde{b}= 20, 22$.

Indeed, we can see explicitly from the form of the Morrison--Park
Weierstrass model \cref{eq:Morrison--Park} that for the cases
$\tilde{b}= 20, 22$, which correspond to choosing classes $[c_3] =
\tilde{b}/2 =10, 11$, the class $[c_0]$ becomes ineffective and there
is an enhancement to an additional $\asu(2)$ factor on the class
$[c_1]$.  This enhancement was also discussed from different points of
view in \cite{MorrisonParkU1, MorrisonTaylorMultisection}.  More
generally, whenever we have a $\au(1)$ theory with generic matter and
an anomaly coefficient $\tilde{b}$ such that $- 6 a - \tilde{b}$ is
outside the positivity cone but $- 8 a - \tilde{b}$ is in the (closure
of the) cone, we have a similar situation where the theory suffers
automatic enhancement to $\au(1) \oplus \asu(2)$.  In the F-theory
construction, it can be confirmed that the global structure of the
group in such situations is $(\U(1) \times\SU(2))/\Z_2$
\cite{MorrisonTaylorCompleteness}.

There are also similar situations when $- 6 a - \tilde{b}= 0$, in
which case the enhancement is to a $\U(1) \times\U(1)$ theory, as can be seen again directly
from the Morrison--Park form, and when $\tilde{b}= -8 a$, which
naively gives a $\au(1)$ model with only even charges, but where the
presence of an additional section in the Morrison--Park model makes
this equivalent to the model with only charges $\chargedsinglet{1}$, in
accord with the Massless Charge Sufficiency Conjecture
\cite{MorrisonTaylorCompleteness}.  Note that the original
Morrison--Park generating section in the enhanced $\U(1) \times\U(1)$
theory when $-6a = \tilde{b}$ is the ``diagonal'' sum of the
two $\U(1)$
generators giving the charge lattice generated by $\twochargedsinglet{\pm1}{0}$ and
$\twochargedsinglet{0}{\pm1}$.
These cases play a further role as a special case in the analysis of
the following subsection.

\subsection{$\U(1)$ infinite families}
\label{sec:abelian-infinite}

We now turn to some families of $\U(1)$ models with exotic charges $q
> 2$.
It was shown in \cite{TaylorTurnerSwampland} that there is an infinite
family of $\U(1)$ models with no tensor multiplets that satisfy the
6D anomaly cancellation conditions and other known low-energy
constraints. The models in this family have $\U(1)$ height
\begin{equation}
    \height = 6\left(q^2 + q r + r^2\right)
\end{equation}
and charged hypermultiplet spectra of the form\footnote{The sign of
  $\U(1)$ charge for a charged hypermultiplet is unimportant, as a
  $\chargedsinglet{Q}$ hypermultiplet contains fields with $\U(1)$ charges
  $+Q$ and $-Q$.}
\begin{equation}
    \label{eq:chargefamilyspectrum}
    54\times\chargedsinglet{q}+54\times\chargedsinglet{r}+54\times\chargedsinglet{(q+r)}
\end{equation}
for $q,r \in \Z$. From now on, we ignore situations where either $q$,
$r$, or $q+r$ is 0; after normalizing charges, these situations
correspond to a relatively uninteresting model known to occur in
F-theory \cite{MorrisonParkU1} whose charged hypermultiplet spectrum
has only 108 $\chargedsinglet{1}$ hypermultiplets.  We also assume
that $q, r$ are relatively prime, so that these charges generate the
full possible charge lattice and therefore satisfy the Massless Charge
Sufficiency Conjecture
\cite{MorrisonTaylorCompleteness}.\footnote{Tarazi and Vafa have made the stronger
conjecture \cite{TaraziVafa-conjecture} that there must always be a
massless field of charge $\chargedsinglet{1}$ in a 6D $\U(1)$
supergravity theory of this type, which would be violated by many of
these theories, but there would still be an infinite family with $q
= 1, r$ prime compatible even with this stronger conjecture.}
Because the
number of topologically distinct elliptically fibered Calabi--Yau threefolds is finite, only a
finite number of these supergravity models can be realized in
F-theory. This family therefore presents an infinite number of
swampland candidates, but there has been no proposal as to which
precise subset of
these models lie in the F-theory swampland, let alone why. And since
the models in this family support arbitrarily large $\U(1)$ charges,
this family is relevant for questions regarding which $\U(1)$
charges occur in F-theory (see, e.g.,\cite{RaghuramTaylorLargeCharge,
Cianci:2018vwv, LeeWeigandAbelian, Collinucci:2019fnh, AndersonGrayOehlmannQuotients}).

However, there is also a model with zero tensor multiplets, a $\U(1)\times\U(1)$ gauge group, and a relatively simple spectrum of charged hypermultiplets:\footnote{A $\twochargedsinglet{Q_a}{Q_b}$ hypermultiplet is equivalent to $\twochargedsinglet{-Q_a}{-Q_b}$ hypermultiplet, as both contain fields with charges $(Q_a,Q_b)$ and $(-Q_a,-Q_b)$.}
\begin{equation}
    \label{eq:u1u1spectrum}
    54 \times \twochargedsinglet{1}{-1} + 54\times \twochargedsinglet{1}{0} + 54\times\twochargedsinglet{0}{1}\,.
\end{equation}
The anomaly coefficient matrix for the $\U(1)\times\U(1)$ gauge group is
\begin{equation}
    \height = \begin{pmatrix} 6&  -3 \\ -3 & 6  \end{pmatrix}\,.
\end{equation}
This model appears repeatedly in this section, and unless otherwise specified, any reference to a $\U(1)\times\U(1)$ model refers to this specific model. Let us refer to the first $\U(1)$ factor as $\U(1)_a$ and the second $\U(1)$ factor as $\U(1)_b$. We are also free to use a different basis $\U(1)_{c}\times \U(1)_{d}$, such that\footnote{This transformation is not valid if $q=0$, but as mentioned above this situation is uninteresting.}
\begin{equation}
\begin{pmatrix}q_{c}\\q_{d}\end{pmatrix} = \begin{pmatrix}q&-r\\ 0&1 \end{pmatrix}\begin{pmatrix}q_{a}\\q_{b}\end{pmatrix}\,.
\end{equation}
The charge spectrum in this new basis is
\begin{equation}
54\times\twochargedsinglet{q+r}{-1} + 54\times\twochargedsinglet{q}{0} + 54\times\twochargedsinglet{r}{1}\,,
\end{equation}
and the anomaly coefficient matrix is
\begin{equation}
\begin{pmatrix} 6\left(q^2 + q r + r^2\right) & -3(q+2r) \\ -3(q+2r) & 6 \end{pmatrix}\,.
\end{equation}

Although $\U(1)_c$ corresponds to the $\U(1)$ seen in the infinite
family, there are still two $\U(1)$ factors in this model. One could imagine obtaining a model in the infinite family by Higgsing $\U(1)_{d}$ in a way that preserves $\U(1)_{c}$. This Higgsing would entail giving a \vev{} to $\twochargedsinglet{0}{1}$ matter in the $\U(1)_{c}\times\U(1)_{d}$ basis, but this matter is not present in the spectrum unless $r$ or $r+q$ is 0. The \conjecture\ would then imply that if we attempt to construct any $\U(1)$ model in the family with $r,q,r+q\neq0$, the model is forced to obtain an extra $\U(1)$, and we end up with the $\U(1)\times\U(1)$ model.

The $\U(1)\times\U(1)$ model has an F-theory realization described in \cref{app:u1u1model}. Geometrically, $\U(1)_{a}$ and $\U(1)_{b}$ correspond to the generating sections $\ratsec{s}_{Q}$ and $\ratsec{s}_{R}$, while $\U(1)_c$ corresponds to the linear combination $q\ratsec{s}_{Q}-r\ratsec{s}_{R}$ under elliptic curve addition. The natural way to test the \conjecture\ is to try constructing the models in the infinite family using the known F-theory $\U(1)$ constructions and to show that we end up with the $\U(1)\times\U(1)$ F-theory model. Because the known $\U(1)$ constructions in F-theory admit only relatively small charges, we will only be able to test the conjecture for small $r$ and $q$. Still, we see clear evidence of the expected enhancement in all the cases we consider. When we try to use these constructions to obtain $\U(1)$ models in the infinite family, the divisor classes of various parameters become either ineffective or trivial. As a result, the Mordell--Weil rank enhances, and the section that seems to support the correct $\U(1)$ charges becomes a combination of $\ratsec{s}_{Q}$ and $\ratsec{s}_{R}$.

\paragraph{$r=q=1$}
When $r$ and $q$ are both 1, the desired spectrum of charged hypermultiplets for the $\U(1)$ model takes the form
\begin{equation}
    \label{eq:u1q2spectrum}
    108\times\chargedsinglet{1} + 54\times\chargedsinglet{2}\,.
\end{equation}
Since the $\U(1)$ model only has matter with charges $q=\pm1$ and
$\pm2$, the Weierstrass model should be in the Morrison--Park form \labelcref{eq:Morrison--Park}. We take the base of the elliptic fibration to be $\bP^2$. For $\tilde{b}=6(q^2+q r + r^2)=18$, the parameters in Morrison--Park form would have divisor classes
\begin{equation}
    \divclass{\mpb} = 6H\,, \quad \divclass{c_3} = 9H\,, \quad \divclass{c_2} = 6H\,, \quad \divclass{c_1} = 3H\,, \quad \divclass{c_0} = 0H\,.
\end{equation}
This is precisely an example of the enhancement structure described in
the last paragraph of the preceding subsection.
The divisor class $\divclass{c_0}$ is trivial, and $c_0$ can freely be taken to be a perfect square. But if $c_0$ is a perfect square, the gauge group enhances to $\U(1)\times\U(1)$ \cite{MorrisonTaylorMultisection}. One can verify the presence of this additional $\U(1)$ by counting moduli and comparing to the expectations from the gravitational anomaly cancellation conditions \cite{RaghuramTaylorTurnerSM}. Additionally, we can map the Morrison--Park form with trivial $\divclass{c_0}$ to the $\U(1)\times\U(1)$ model discussed in \cref{app:u1u1model} by relating the parameters as follows:
\begin{equation}
\begin{gathered}
    \mpb =  S_ 7\,, \quad c_0 = \frac{S_ 5^2}{4}\,, \quad c_1 = S_9 S_1-\frac{1}{2} S_ 5 S_ 6\,,\\
    c_2= \frac{1}{4} \left(-4 S_9 S_ 2+2 S_ 5 S_ 7+S_ 6^2\right)\,, \quad c_3= 2 S_9 S_ 3-\frac{1}{2} S_ 6 S_7\,.
\end{gathered}
\end{equation}
Because $\divclass{c_0}$, $\divclass{S_5}$, and $\divclass{S_9}$ are
trivial, this is simply a redefinition of parameters, not a
tuning. Thus, our attempt to obtain the desired $\U(1)$ model with the
Morrison--Park form has recovered the $\U(1)\times\U(1)$ model, as
expected. One can also verify that the standard Morrison--Park section
\begin{equation}
[\secx:\secy:\secz] = \left[c_3^2-\frac{2 \mpb^2 c_2}{3}:-\frac{1}{2} \mpb^4 c_1+\mpb^2 c_2 c_3-c_3^3:\mpb\right]
\end{equation}
is given by $\ratsec{s}_{Q}-\ratsec{s}_{R}$.

\paragraph{$r=2$, $q=1$}
When $r=2$ and $q=1$ (or equivalently $q=2$ and $r=1$), the charged hypermultiplet spectrum takes the form
\begin{equation}
    54\times\chargedsinglet{1} + 54\times\chargedsinglet{2} +54\times\chargedsinglet{3}\,,
\end{equation}
We can try to construct this charge-3 $\U(1)$ model with a blown-down version of the $F_3$ toric hypersurface construction from \cite{KleversEtAlToric}.\footnote{Using the more general charge-3 model from \cite{Raghuram34} does not change the results.} While there is a Weierstrass model for this construction, it is easier to work with an alternative form of the elliptic fibration where the fiber is a cubic curve in a $\bP^2$ ambient space with coordinates $[u:v:w]$:
\begin{equation}
u\left(s_1 u^2 + s_2 u v + s_3 v^2 + s_5 u w + s_6 v w + s_8 w^2\right) + v \left(s_4 v^2 + s_7 v w + s_9 w^2\right) = 0\,.
\end{equation}
We take the base of the elliptic fibration to be $\bP^2$. The charged hypermultiplet spectrum of this construction is
\begin{equation}
\begin{aligned}
    \left(12\canonclass^2-\canonclass(8\divclass{s_7}-\divclass{s_9}) -4 \divclass{s_7}^2 +  \divclass{s_7}\divclass{s_9}-\divclass{s_9}^2\right)&\times\chargedsinglet{1} \\
    + \left(6\canonclass^2-\canonclass(4\divclass{s_9}-5\divclass{s_7}) + \divclass{s_7}^2 + 2 \divclass{s_7}\divclass{s_9}-2\divclass{s_9}^2\right)&\times\chargedsinglet{2} \\
    + [s_9]\cdot\left(-\canonclass+\divclass{s_9} -\divclass{s_7}\right)&\times\chargedsinglet{3}\,,
\end{aligned}
\end{equation}
and the anomaly coefficient is $\tilde{b}=-6\canonclass-2\divclass{s_7}+4\divclass{s_9}$.  There are two combinations of $(\divclass{s_7},\divclass{s_9})$ that would seem to give the correct charged hypermultiplet spectrum: $(6H, 9H)$ and $(0H,6H)$.

This model is in fact equivalent to the $\U(1)\times\U(1)$ F-theory model for either choice of $(\divclass{s_7},\divclass{s_9})$. If we focus on the $(\divclass{s_7},\divclass{s_9})=(0H,6H)$ possibility, the divisor classes of the other parameters are
\begin{equation}
    \divclass{s_1} = 3H\,, \quad \divclass{s_2} = 0\,, \quad \divclass{s_3} = -3H\,, \quad \divclass{s_4} = -6H\,, \quad \divclass{s_5} = 6H\,, \quad \divclass{s_6} = 3H\,, \quad \divclass{s_8} =  9H\,.
\end{equation}
Because $\divclass{s_3}$ and $\divclass{s_4}$ are ineffective, $s_3$ and $s_4$ can be set to 0, and the elliptic fibration takes the form
\begin{equation}
u\left(s_1 u^2 + s_2 u v + s_5 u w + s_6 v w + s_8 w^2\right) + v w\left(s_7 v + s_9 w\right) = 0\,.
\end{equation}
If we then exchange $v\leftrightarrow w$ and identify
\begin{equation}
\begin{gathered}
    s_9 = S_7\,, \quad s_7 = S_9\,, \quad  s_2 = S_5\,, \quad s_5 = S_2\,, \\
    s_8 = S_3\,, \quad s_1 = S_1\,, \quad s_6 = S_6\,,
\end{gathered}
\end{equation}
we recover \cref{eq:u1u1p2desc}, which describes the $\U(1)\times\U(1)$ model. Additionally, one can show that the section supporting the $r=2$, $q=1$ spectrum is given by $2\ratsec{s}_{Q}-\ratsec{s}_{R}$. A similar story holds for $(\divclass{s_7},\divclass{s_9})=(6H,9H)$. We therefore see the expected enhancement to the $\U(1)\times\U(1)$ model for $r=2$ and $q=1$.

\paragraph{$r=3$, $q=1$} When $r=3$ and $q=1$ (or when $q=3$ and $r=1$), the $\U(1)$ charged hypermultiplet spectrum takes the form
\begin{equation}
    54\times\chargedsinglet{1} + 54\times\chargedsinglet{3} +54\times\chargedsinglet{4}\,.
\end{equation}
We can attempt to construct this model using the charge-4 $\U(1)$ Weierstrass model in \cite{Raghuram34}. According to the matter spectrum given in \cite{Raghuram34}, we should let $\divclass{a_1}$ be $6H$, $\divclass{b_1}$ be $9H$, and $\divclass{d_1}$ be $0H$ to obtain the desired $\U(1)$ spectrum.\footnote{We could also let $\divclass{a_1}=9H$ and $\divclass{b_1}=6H$, but this situation is equivalent to the $\divclass{a_1}=6H, \divclass{b_1}=9H$ model.} The other parameters have the following divisor classes:
\begin{equation}
\begin{aligned}
\divclass{d_0}&=-3H & \divclass{d_2}&=3H  &  \divclass{s_1}&=-6H & \divclass{s_2}&=-3H \\
\divclass{s_3}&=0H  & \divclass{s_5}&=0H & \divclass{s_6}&=3H & \divclass{s_8} &= 6H
\end{aligned}
\end{equation}
Because their divisor classes are ineffective, the parameters $d_0$, $s_1$, and $s_2$ should be set to 0, giving us a Weierstrass model described by
\begin{equation}
\begin{aligned}
    f &= -\frac{1}{48} \left(s_6^2-4 s_3 s_8\right)^2-\frac{1}{2} s_5 s_6 s_3 \left(a_1 d_2+b_1 d_1\right) \\
    &\qquad\qquad - \frac{1}{3} a_1^2 d_1^2 s_5^2+\frac{1}{6} a_1 d_1 s_5 \left(s_6^2+2 s_3 s_8\right)+b_1 d_2 s_5 s_3^2
\end{aligned}
\end{equation}
and
\begin{equation}
\begin{aligned}
    g &= \frac{1}{864} \left(s_6^2-4 s_3 s_8\right)^3+\frac{1}{4} s_3^2 s_5^2 \left(a_1^2 d_2^2+b_1^2 d_1^2\right)-\frac{1}{6} s_3 s_6 s_5^2 \left(a_1 b_1 d_1^2+a_1^2 d_1 d_2\right) \\
&\qquad\qquad + \frac{1}{24} s_3 s_6 \left(s_6^2-4 s_3 s_8\right) s_5 \left(a_1 d_2+b_1 d_1\right) \\
&\qquad\qquad - \frac{1}{72} a_1 d_1 \left(12 b_1 d_2 s_3^2 s_5^2+\left(s_6^2-4 s_3 s_8\right) \left(s_6^2+2 s_3 s_8\right) s_5\right) \\
&\qquad\qquad - \frac{2}{27} a_1^3 d_1^3 s_5^3+\frac{1}{18} a_1^2 d_1^2 \left(s_6^2+2 s_3 s_8\right) s_5^2-\frac{1}{12} b_1 d_2 s_3^2 \left(s_6^2-4 s_3 s_8\right) s_5\,.
\end{aligned}
\end{equation}
This Weierstrass model is in fact equivalent to the $\U(1)\times\U(1)$
Weierstrass model of \cref{app:u1u1model}. To see this, we first note
that we can remove all occurrences of $s_5$ through the rescalings
$d_2\to s_5^{-1} d_2$ and $d_1\to s_5^{-1} d_1$;  since
$\divclass{s_5}$ is trivial, the divisions by $s_5$ do not cause
issues. Then, we can identify the parameters in the two Weierstrass models as follows:
\begin{equation}
S_1 = d_2\,, \quad S_2 = s_8\,, \quad S_3 = b_1\,, \quad S_5 =d_1\,, \quad S_6 = s_6\,, \quad S_7 = a_1\,, \quad S_9 = s_3\,.
\end{equation}
The Weierstrass models are equivalent, implying that the gauge group is $\U(1)\times\U(1)$ instead of just $\U(1)$. And one can verify that the section that seems to support the $r=3, q=1$ spectrum is in fact equivalent to $3\ratsec{s}_{Q}-\ratsec{s}_{R}$. We therefore see the expected automatic enhancement when we attempt to construct the $r=3$, $q=1$ $\U(1)$ model.

A similar infinite family of $\U(1)$ models with one tensor multiplet was given in \cite{ParkTaylorAbelian}. The charged hypermultiplet spectra of these models, which take the form
\begin{equation}
    \label{eq:chargefamilyspectrumT1}
    48\times\chargedsinglet{r} + 48\times\chargedsinglet{s} + 48\times\chargedsinglet{(r+s)}\,,
\end{equation}
mimic the structure of the $T=0$ family discussed above. The analysis in \cite{ParkTaylorAbelian} pointed out that there is a related $\U(1)\times\U(1)$ model whose charged hypermultiplet spectrum is
\begin{equation}
    48\times\twochargedsinglet{1}{0} + 48\times\twochargedsinglet{0}{1} + 48\times\twochargedsinglet{1}{1}\,.
\end{equation}
As with the $T=0$ infinite family, one can view the $\U(1)$ algebras
in the $T=1$ infinite family as linear combinations of the two $\U(1)$
algebras in this $\U(1)\times\U(1)$ model. But we lack the matter
necessary to properly Higgs the $\U(1)\times\U(1)$ model down to the
desired $\U(1)$ models unless $r$,$s$, or $r+s$ is 0. The
\conjecture\ therefore implies that if we attempt to construct the
models in this infinite family, the gauge group should enhance to
$\U(1)\times\U(1)$. One can confirm that this occurs in the
$(r,s)=(1,1)$, $(2,1)$, and $(3,1)$ cases by an explicit analysis of F-theory constructions along the lines above; some aspects of this analysis are discussed in \cref{app:u1u1test}.

There is another infinite family of $\U(1)$ models \cite{TaylorTurnerSwampland}  with zero tensor multiplets and charged hypermultiplet spectra of the form
\begin{equation}
    \label{eq:fourchargefamily}
    54 \times \chargedsinglet{a} + 54 \times \chargedsinglet{b} + 54 \times \chargedsinglet{c} + 54 \times \chargedsinglet{d}\,,
\end{equation}
where
\begin{equation}
    a = m^2 - 2 m n\,, \quad b = 2 m n-n^2\,, \quad c = m^2 - n^2\,, \quad d = 2(m^2 - m n + n^2)
\end{equation}
for $m,n \in \Z_{+}$ with $n\le \frac{m}{2}$. Some of these models belong to the three-charge infinite family of \cref{eq:chargefamilyspectrum}, suggesting that the \conjecture\ may be important for the four-charge family as well. However, we have not yet found an enhanced model for the nontrivial four-charge models. Part of the difficulty lies in the complicated nature of the models: because even the simplest non-trivial members of the family have $\U(1)$ charges larger than those found in the known F-theory constructions, we cannot directly observe any enhancements. And  the quadratic structure of the charges, exhibited by their dependence on $m$ and $n$, contrasts with the linear nature of \cref{eq:chargefamilyspectrum,eq:chargefamilyspectrumT1} that naturally reflects combinations of multiple $\au(1)$ algebras. In fact, a similar quadratic structure appears in the $\asu(13)\oplus\au(1)$ family of \cref{sec:13}. Understanding whether automatic enhancement plays a role in this four-charge family is an important direction for future work.

To summarize, the \conjecture\ implies that none of the $\U(1)$ models in the infinite families of \cref{eq:chargefamilyspectrum,eq:chargefamilyspectrumT1} (except for trivial cases with only one type of non-zero charge) can be realized in F-theory, as the gauge group is forced to enhance to $\U(1)\times\U(1)$. The conjecture thus offers a reason as to why those models in the infinite family lie in the F-theory swampland. While finding the ``failure mode'' for this family is an important step, it does not definitely tell us which $\U(1)$ charges can generally occur in F-theory. Yet this result offers a valuable lesson: limits on the maximum $\U(1)$ charge in F-theory likely come not from local constraints on the behavior of sections but from global considerations \cite{MorrisonParkU1}. The problem with models in the infinite family lies in an enhancement of the Mordell--Weil rank, which is a global property of the fibration.

\subsection{Infinite family of $\asu(13)\oplus\au(1)$ models}
\label{sec:13}

In \cite{ParkTaylorAbelian}, an infinite family of $T=1$ supergravity models with $\asu(13)\oplus\au(1)$ gauge algebras was presented. The charged hypermultiplet spectra for these models take the form
\begin{equation}
    \label{eq:su13familyspec}
    \bm{78}_{-3a-\frac{2}{3}f} + 3\times\bm{78}_{a} + 6\times\bm{13}_f + 18\times\bm{1}_{6a+f}\,,
\end{equation}
where
\begin{equation}
\begin{aligned}
    a&= 13 r^2 - 234 r s - 51 s^2 & f&= 24(13 r^2 +3 s^2)\\
    r&= 84 n+ 43 & s&= 182 n + 92
\end{aligned}
\end{equation}
for $n \in \Z$. The anomaly coefficient for the $\au(1)$ algebra is given by $\tilde{b}=\frac{1}{2}(\alpha,\tilde{\alpha})$ with
\begin{equation}
\begin{aligned}
    \alpha &= 52\left[6687s^4+ 54756 s^3 r + 94458 s^2 r^2 - 124956 s r^3 + 39455 r^4 \right]\,,\\
    \tilde{\alpha} &= 52\left[2475s^4 + 37908 s^3 r + 170274 s^2 r^2 - 29484 s r^3 + 9035r^4\right]\,,
\end{aligned}
\end{equation}
in a basis where the intersection form is
\begin{equation}
    \Omega = \begin{pmatrix} 0 & 1 \\ 1 & 0 \end{pmatrix}\,.
\end{equation}

First, note that if we attempt to obtain just the $\asu(13)$ algebra, the model suffers an automatic enhancement. In order to obtain four antisymmetric and six fundamental hypermultiplets of $\asu(13)$, the anomaly coefficient $b$ for the $\asu(13)$ algebra should satisfy $b\cdot b = 2$ and $-a\cdot b=4$, implying that $(-6a-13b)\cdot b$ is negative and that $-6a-13b$ is ineffective. By the discussion in \cref{sec:t0-general-n},
the $\asu(13)$ algebra should automatically enhance to $\asu(13)\oplus\au(1)_f$, where we use the subscript $f$ to distinguish this $\au(1)$ algebra from those in the infinite family.\footnote{Note that $-3a-6b$ is effective, and there should not be automatic enhancements beyond the one to $\asu(13)\oplus\au(1)$.} The charged hypermultiplet spectrum
\begin{equation}
    \label{eq:su13autoenhancespec}
    4\times\bm{78}_{-\frac{1}{13}} + 6\times\bm{13}_{\frac{6}{13}}
\end{equation}
lacks the necessary charged singlets to Higgs away the forced $\au(1)$ algebra, in line with the \conjecture.

Strictly speaking, this statement only applies to a model with an $\asu(13)$ gauge algebra, not the $\asu(13)\oplus\au(1)$ models in the infinite family. But because we are able to phrase this automatic enhancement locally in terms of $a\cdot b$ and $b\cdot b$, we would expect the $\asu(13)$ subalgebra of $\asu(13)\oplus\au(1)$ to also experience an automatic enhancement. A naive guess is that the gauge algebra for the infinite family automatically enhance to $\asu(13)\oplus\au(1)\oplus\au(1)_f$, but if one combines the $\au(1)$ charges in \cref{eq:su13familyspec} and \cref{eq:su13autoenhancespec}, the resulting spectrum does not satisfy the anomaly cancellation conditions. Nevertheless, the expected automatic enhancement of the $\asu(13)$ algebra suggests models in this $\asu(13)\oplus\au(1)$ infinite family should have some inconsistency, even though we have not found an explicit automatic enhancement.

This argument does not seem to capture the true problem with the
family, however. There are also seemingly consistent
$\asu(N)\oplus\au(1)$ infinite families for $N<13$, which can be found
by Higgsing the would-be $\asu(13)\oplus\au(1)$ models on
fundamental matter. The $\asu(N)$ subalgebras would not suffer the
automatic enhancement of $\asu(13)$ described above, but the
mechanism that eliminates the $N<13$ families would presumably apply
to the $\asu(13)\oplus\au(1)$ family as well. We therefore expect
there to be an alternative, more satisfying explanation that rules
them out. The quadratic nature of the $\au(1)$ charges resembles
that seen in the four-charge family of \cref{eq:fourchargefamily},
suggesting that all of these families suffer from similar
issues. But it is difficult to see how the quadratic structure of the charges
would fit naturally in any kind of linear $\au(1)$ subalgebra of an
enhanced $\U(1)^k$ theory. Additionally, the charges are significantly larger than
those that can be easily realized in F-theory constructions, so it
would be tough to directly observe how the models fail.
It would be interesting to understand these infinite families more properly in future work.

\subsection{Automatic enhancement for $(\SU(3) \times \SU(2) \times
\U(1)) / \Z_6$ models}
\label{sec:SM}

In \cite{RaghuramTaylorTurnerSM}, the universal Weierstrass model was
constructed for the Standard Model gauge group $(\SU(3) \times \SU(2)
\times \U(1)) / \Z_6$. As discussed there, the model undergoes
automatic enhancement for certain choices of the anomaly coefficients
$b_3, b_2, \beta$. In particular, the Jacobian rank method introduced
in that paper was used to count the number of moduli for all 98 of the
allowed parameter choices for 6D models over the base
$\bP^2$. In 44 of these cases the moduli count does not
match with the expectation from the spectrum fixed by anomaly
cancellation, and the model undergoes automatic enhancement or is
rendered inconsistent.

In 34 of these cases, the gauge group enhances by the addition of
nonabelian or $\U(1)$ factors. In each of these cases, the resulting
unHiggsed theory is anomaly consistent but has insufficient matter to
supersymmetrically Higgs to the original $(\SU(3) \times \SU(2) \times
\U(1)) / \Z_6$ model, consistent with the \conjecture. In the
remaining 10 cases, the discriminant vanishes identically, indicating
that the Weierstrass model is invalid. These are all models that would
apparently have an unHiggsing to a Tate $\SU(4) \times \SU(3) \times
\SU(2)$ model.
However, it is found that the corresponding Tate models
also have identically vanishing discriminant, corresponding to the
stronger case-specific version of the enhancement conjecture that when
certain positivity conditions are violated the theory is either
enhanced or inconsistent. The various cases are summarized in
\cref{tab:321AEC}.

\begin{table}
    \centering

    \[\begin{array}{*{11}{c}} \toprule
        \divclass{b_3} & \divclass{d_0} & \divclass{s_1} & \divclass{s_2} & \divclass{d_1} & \divclass{d_2} & \divclass{s_5} 
            & \divclass{s_8} & \text{\# of Models} & \text{Gauge Algebra} \\ \midrule
        +              & +              & \ge            & -              & \ge            & +              & +              
            & \ge            & 11                  & \asu(3) \oplus \asu(2) \oplus \au(1) \oplus \au(1) \\
        +              & +              & -              & -              & \ge            & +              & \ge            
            & 0              & 9                   & \asu(3) \oplus \asu(3) \oplus \au(1) \\
        +              & +              & -              & -              & +              & +              & +              
            & +              & 4                   & \asu(3) \oplus \asu(3) \oplus \asu(2) \oplus \au(1) \\
        +              & +              & 0              & -              & +              & +              & \ge            
            & -              & 3                   & \asu(4) \oplus \asu(2) \oplus \au(1) \oplus \au(1) \\
        +              & +              & \ge            & -              & 0              & +              & 0              
            & 0              & 3                   & \asu(3) \oplus \asu(2) \oplus \au(1) \oplus \au(1) \oplus \au(1) \oplus \au(1) \\
        +              & +              & 0              & -              & 0              & +              & -              
            & -              & 1                   & \asu(5) \oplus \asu(2) \oplus \au(1) \oplus \au(1) \\
        +              & +              & +              & -              & +              & +              & 0              
            & -              & 1                   & \asu(4) \oplus \asu(2) \oplus \asu(2) \oplus \au(1) \oplus \au(1) \\
        +              & +              & 0              & -              & +              & +              & -              
            & -              & 1                   & \asu(5) \oplus \asu(2) \oplus \asu(2) \oplus \au(1) \oplus \au(1) \\
        +              & +              & +              & -              & 0              & +              & -              
            & -              & 1                   & \asu(5) \oplus \asu(3) \oplus \asu(2) \oplus \au(1) \oplus \au(1) \\
        +              & +              & -              & -              & \ge            & +              & \pm            
             & -              & 10                  & \text{Invalid} \\ \bottomrule
    \end{array}\]

    \caption{Cases where the $(\SU(3) \times \SU(2) \times \U(1)) / \Z_6$ universal Weierstrass model undergoes automatic enhancement over the base $\bP^2$. Cases are grouped by their resulting gauge algebra, which is determined by which divisor classes $[b_3]\text{--}[s_8]$ associated with parameters in the Weierstrass model are strictly effective ($+$), effective ($\ge$), trivial ($0$), ineffective ($-$), or unspecified ($\pm$). The parameter $s_6$ is not listed and always has class $-\canonclass = 3 H$.}
    \label{tab:321AEC}
\end{table}

\section{Positivity conditions and
local conditions for automatic enhancement}
\label{sec:local}

In each specific class of cases we have encountered in the preceding
sections, the pattern is essentially that when we try to tune an
F-theory model with algebra $\aG$ with a certain anomaly coefficient
$b$ (or $\tilde{b}$), and when the curve class associated with
a certain other string charge $-n a-mb$ ceases to be effective,
then either an additional gauge factor with algebra of at least $\aF$
arises with another anomaly coefficient $b'= -pa-qb$, or the theory
becomes inconsistent due to excessive singularities.
Put more simply, under such circumstances we cannot have a theory with
gauge algebra $\aG$ having anomaly coefficient $b$ satisfying the
stated conditions unless there is also a gauge algebra component of at
least $\aF$
associated with the anomaly coefficient $b'$.

In the context of F-theory, for each such algebra enhancement, the
conjecture can thus be stated in terms of the effectiveness of certain
curves; in supergravity this translates to conditions that certain
anomaly coefficients or other elements of the string charge lattice
lie in or outside the positivity cone.
While we do not have any completely general way of unifying these
positivity conditions into a single conjecture, in each individual
case, the statement about positivity conditions is stronger than the
\conjecture, since it includes those cases where the theory is
rendered invalid by the would-be enhancement.  In this section we collect the
specific results found in the other sections for this stronger condition.

We also collect here some  corresponding local versions of the
\conjecture.  In each case,  the general form of the statement is that
if there is a gauge component $\aG$ with anomaly coefficient $b$ satisfying
certain conditions, then a consistent F-theory model is only possible
when there is an additional gauge component
of at least $\aF$ associated with another
anomaly coefficient $b'$.
The conditions in the local form for each case only depend on $b \cdot a, b \cdot b$, which
correspond to local conditions in the F-theory geometry, namely the
genus and self-intersection of the associated curve.
These conditions are weaker than the global positivity conditions, but
are useful in analyzing models using only part of the structure of the theory.

A number of these conditions that we have
encountered in the paper are listed in \cref{tab:local}.

\begin{table}
    \centering

    \renewcommand{\arraystretch}{1.2}
    \[\begin{array}{*{5}{c}}
    \toprule
    \text{Algebra } \aG &  \text{Positivity Condition} & \text{Local Condition} & \aF & b'
    \\
    \midrule
    \text{---} & \text{---} & b \cdot b < -2, g = 0 & G' \supseteq \SU(3) & b \\
    \text{---} & \text{---} & b \cdot b \le -13, g = 0 &  \text{no theory} & \text{---} \\
    \text{---} & \text{---} & b \cdot b < -2, g > 0 & \text{no theory} & \text{---} \\
    \midrule
    \asu(2) & -3a-b  >0 &
    b \cdot b > 0, b \cdot b > -3a \cdot b &
    \asu(2) & -4a-b \\
    \asu(2) & b  = -3a & \text{---} &
    \au(1) & \tilde{b}' =-8a-2b \\
    \asu(3) &-2a-b > 0  & b \cdot b > 0, b \cdot b > -2a \cdot b &
    \asu(3) & -3a-b \\
    \asu(3) &  b = -2a & \text{---} &
    \au(1)\oplus \au(1) &
    {\renewcommand{\arraystretch}{1} \tilde{b}' =
    -a \text{\small $\begin{pmatrix} 6&-3\\-3&6\end{pmatrix}$}} \\
    \asu(4) & -3a-2b > 0  &b \cdot b > 0, 2b \cdot b > -3a \cdot b &
    \asu(2) & -4a-2b \\
    \asu(n=\text{20--23}) & \text{---} & g = 0, b \cdot b = 1 &
    \asu(2) &  -4a-\lceil n/2\rceil b \\
    \au(1) & -6a-\tilde{b} > 0  &
    \tilde{b} \cdot \tilde{b} > 0, \tilde{b} \cdot \tilde{b} > -6a \cdot \tilde{b} &
    \asu(2) &   -4a-\tilde{b}/2 \\
    \au(1) &  \tilde{b} = -6a & \text{---} &
    \au(1) & {\renewcommand{\arraystretch}{1} \tilde{b}' =
    -a \text{\small $\begin{pmatrix} 2&-1\\-1&2\end{pmatrix}$}} \\
    \bottomrule
    \end{array}\]

    \caption{Positivity cone and
    local versions of the \conjecture, for cases
      with generic matter.
    In each case a gauge component $\aG$ with anomaly coefficient $b$
    satisfying certain conditions forces enhancement to an additional
    gauge component of at least
    $\aF$ with anomaly coefficient  $b'$.  (For $\au(1)$ factors, anomaly
    coefficients  are $\tilde{b}, \tilde{b}'$ respectively; note that this
    anomaly coefficient corresponds to the normalization of the $\au(1)$
    factor so that all charges are integers.)
    The ``genus'' contribution $g$
    is given in terms of local anomaly coefficients $a, b$ by \cref{eq:genus}.}
    \label{tab:local}
\end{table}

\noindent Some comments on these local versions of the \conjecture:
\begin{itemize}
    \item Unlike the general statement of the \conjecture, the local
    versions of these statements are specific to rather particular
    circumstances of gauge groups and anomaly coefficients.  The list here
    is not in any way intended to be complete, and even for those cases
    listed, we have given only some simple sufficient local conditions for each
    enhancement.  It would
    also be desirable to have a more unified way of framing these in terms
    of a more general statement analogous to \conjecture.

    \item In cases where $b'$ is not in the positive cone, there is no
    corresponding F-theory model --- the enhancement breaks the theory.

    \item The local conditions given here are only sufficient
    conditions, since there can be an enhancement that depends on further
    structure. In particular, even when the sufficient local conditions
    are not satisfied, the presence of certain other types of curves
    intersecting the curve supporting the gauge divisor can provide for an
    enhancement in more general circumstances.  An example of this is
    given in, for example, \cref{sec:small-bb}, where we see that the
    usual enhancement that begins at $\asu(13)$ on a curve of genus 0 and
    self-intersection $b \cdot b = 2$ also occurs for $\asu(10$--$12)$ when
    the curve in question intersects another curve $b'$ of
    self-intersection $b' \cdot b' = -1$.  A more careful and thorough
    analysis of local conditions incorporating, e.g., adjacent divisors
    could be made using similar principles to those used in
    \cite{JohnsonTaylorEnhanced}.

    \item As for the general \conjecture, the positivity based and
    local versions given here for specific gauge factors are mostly
    conjectural even at the level of F-theory. The case of non-Higgsable
    clusters $(G = 1)$ was proven in full generality in F-theory in
    \cite{MorrisonTaylorClusters}, but the other examples  depend upon some assumptions on the form of the Weierstrass model
    used to achieve the initial gauge algebra $G$.  In particular, while
    the general form
    has been proven for, e.g.,the $\asu(n)$
    algebras under the UFD assumption \cite{MorrisonTaylorMaS}, it
    is possible that more general forms of the Weierstrass model such as
    those considered in \cite{KleversEtAlExotic} may evade the conjecture
    at both the local and global levels.  Although such non-UFD constructions
    in general are expected to give exotic matter representations, this
    has not been proven to always be the case. Nonetheless, as discussed
    in \cref{app:u1u1test}, a variety of effects can remove non-UFD
    structure, and based on the example presented there it seems likely that these effects conspire to ensure that models without exotic matter take UFD forms.

    $\bullet$
    As for the general form of the
    \conjecture, it is natural to speculate that the  group-specific
    stronger versions of
    the conjecture also hold for 6D supergravity theories more generally;
    this possibility is discussed further in the conclusions section.

    \item We have only listed here cases with generic matter, which
    depend in general on knowing the corresponding universal Weierstrass
    models.

    \item As mentioned earlier,
    in cases where the automatic enhancement involved would
    break the theory, such as by producing codimension one (4, 6) loci or
    when $b'$ is not in the positive cone,
    the  gauge-specific positivity-based form and the
    local form of the \conjecture\ still effectively rule out the
    relevant swampland models, while the global form as stated in
    \cref{sec:conjecture} does not due to the absence of a consistent
    containing theory.  Several examples of this are given in
    \cref{sec:break,sec:13}.

    \item In each case where  the local enhancement mechanism forces a
    gauge group factor with a new anomaly coefficient $b'$, as discussed
    in \cref{sec:su2su2}, this can also be achieved by having an
    enhanced gauge factor on each of a set of anomaly coefficients that
    add to $b' = b_1+ \cdots b_k$.
\end{itemize}

\section{Exotic matter and discrete gauge groups}
\label{sec:exotic-discrete}

In this section we discuss several further classes of apparent
swampland models, those involving exotic matter and those involving
discrete gauge groups.  We first describe some cases of exotic matter
types that should be allowed in F-theory but where certain
combinations of these matter representations lie in the apparent
swampland.  We then briefly describe exotic matter that goes outside
the F-theory classification of \cite{KleversEtAlToric} and situations
with discrete gauge groups.  These latter two classes of examples are
still poorly understood in F-theory, and for these types of apparent
swampland models some further work and insights may be needed to fit
with the context of the automatic enhancement mechanisms discussed in
this paper.

\subsection{An $\asu(2)$ model with two half-hypermultiplets of
  $4$ matter.}
\label{sec:exotic-F-theory}

In addition to the fundamental and adjoint representations of $\asu(2)$, F-theory models can realize matter in the the three-index symmetric ($\bm{4}$) representation. Such matter is supported at triple point singularities of the divisor supporting the $\asu(2)$ algebra.  More precisely, a triple point singularity of an $\asu(2)$ divisor in a 6D F-theory model may support a half-hypermultiplet of $\bm{4}$ matter and a full hypermultiplet of fundamental ($\bm{2}$) matter \cite{KleversTaylorTriple, KleversEtAlExotic}.\footnote{It is also possible for triple points to support adjoint matter \cite{KleversEtAlExotic}, but this possibility is not important for the models considered here.}  While there are several matter spectra with $\bm{4}$ matter that occur in F-theory models, there are also matter spectra that are consistent with the anomaly conditions but pose problems when one tries to construct them in F-theory. One example, described in \cite{KleversEtAlExotic}, is a model with no tensor multiplets, an $\asu(2)$ gauge algebra, and the following spectrum of hypermultiplets:
\begin{equation}
\label{eq:quinticsu2spec}
2\times\frac{1}{2}\bm{4} + 84\times\bm{2} + 104\times\bm{1}\,.
\end{equation}
This spectrum satisfies the anomaly cancellation conditions with $\gaugedivclass = 5$. We would therefore expect the corresponding F-theory model to have a $\bP^2$ base, and the $\asu(2)$ algebra should be tuned on a quintic curve. Because the model supports two half-hypermultiplets of $\bm{4}$ matter, the quintic curve should have two triple points.

But as pointed out in \cite{KleversEtAlExotic}, an irreducible quintic curve on $\bP^2$ cannot have two triple point singularities. If such a quintic existed, we could always find a line that goes through both of the triple points. According to Bezout's theorem, this line should intersect the quintic at five points, counted with multiplicity. The line and the quintic intersect at the two triple points, and each of these intersections would have multiplicity three. Therefore, the triple points alone would give us an intersection number of $6$, which is greater than the expected value of $5$ for the total intersection number from Bezout's theorem. This contradiction shows that an irreducible quintic curve cannot have two triple points.

There is a potential loophole in this argument: the quintic curve can contain the line passing through the triple points. In this case, the quintic curve would be reducible. For instance, suppose that, without loss of generality, the two distinct triple points are described by the intersection of a line $\locus{\trplb=0}$ and a quadratic curve $\locus{\trpla=0}$.
A quintic curve of the form
\begin{equation}
\label{eq:reduciblequintic}
\trplb\left(3\tb \trpla^2 + 3\tc \trpla \trplb + \td \trplb^2\right)=0
\end{equation}
with
\begin{equation}
    \divclass{\tb} = 0H\,, \quad \divclass{\tc} = 1H\,, \quad \divclass{\td} = 2H
\end{equation}
would then appear to have two triple points. Of course, this quintic
curve reduces to the union of a line and a quartic curve. To deform
this reducible quintic into an irreducible quintic with two triple
points, we would need to introduce a term of the form $\ta
\trpla^3$.
(All of the terms in the quintic must be polynomials of order three in $\trpla,\trplb$ to
preserve the triple point singularities of the quintic and the non-UFD $\asu(2)$ structure \cite{KleversEtAlExotic}.)
 The parameter $\ta$ would then be ineffective and would be
forced to vanish, giving us the reducible quintic of
\cref{eq:reduciblequintic} once again. Clearly, there is a geometric
obstruction to deforming the reducible quintic into an irreducible
curve.

If we were to tune $I_2$ singularities along the curve in \cref{eq:reduciblequintic}, the resulting gauge algebra would be $\asu(2)\oplus\asu(2)$ rather than $\asu(2)$. We take the first $\asu(2)$ to correspond to the line $\locus{\trplb=0}$ and the second $\asu(2)$ to correspond to the quartic curve. One can find the corresponding Weierstrass model by setting $\ta$ to 0 in the $\asu(2)$ construction given in Appendix B of \cite{KleversEtAlExotic}. Each $\locus{\trpla=\trplb=0}$ point supports a half-hypermultiplet of $(\bm{2},\bm{3})$ matter and a half-hypermultiplet of $(\bm{2},\bm{1})$ matter. The complete hypermultiplet spectrum for this model is
\begin{equation}
\label{eq:quinticsu2su2spec}
2\times\frac{1}{2}(\bm{2},\bm{3})  + 1\times(\bm{1},\bm{3}) + 19\times(\bm{2},\bm{1}) + 64\times(\bm{1},\bm{2}) + 104\times(\bm{1},\bm{1})\,.
\end{equation}

The appearance of the $\asu(2)\oplus\asu(2)$ algebra seems linked to
the ineffectiveness of the parameter $\ta$. This is reminiscent of the
automatic enhancement examples encountered above, in which the gauge
algebra enhances because a parameter becomes ineffective. It is
therefore tempting to view the forced reducibility of the quintic
curve as a geometric manifestation of the \conjecture. Specifically,
we argue that the conjecture implies that any attempt to construct an
$\asu(2)$ model with $\gaugedivclass = 5$ and two half-hypermultiplets
of $\bm{4}$ matter leads to a model with an $\asu(2)\oplus\asu(2)$
gauge algebra. In order to show that this is the case, we must examine how one would field-theoretically break the $\asu(2)\oplus\asu(2)$ algebra down to $\asu(2)$.

We are attempting to recombine the 7-branes wrapped on the linear and quartic curves of \cref{eq:reduciblequintic} while maintaining the triple points at $\locus{\trpla=\trplb=0}$. Field-theoretically, this recombination process involves giving a \vev{} to bifundamental matter such that a diagonal $\asu(2)$ subalgebra of $\asu(2)\oplus\asu(2)$ is preserved. Specifically, if the generators for the $\asu(2)$ on the quartic curve are $T^a_{(4)}$ and the generators for the $\asu(2)$ on the linear curve are $T^a_{(1)}$, we wish to preserve the generators
\begin{equation}
T^a_{(5)} = T^a_{(4)} + T^a_{(1)}
\end{equation}
up to some normalization. Of course, the hypermultiplet spectrum in \cref{eq:quinticsu2su2spec} does not contain any bifundamental matter, so we cannot perform this Higgsing process. But if we ignore this problem, it naively seems that this Higgsing would give us the $\asu(2)$ hypermultiplet spectrum in \cref{eq:quinticsu2spec}. The $\asu(2)\oplus\asu(2)$ representations in \cref{eq:quinticsu2su2spec} branch to representations of the diagonal $\asu(2)$ as follows:
\begin{equation}
\label{eq:quinticbranching}
(\bm{2},\bm{3}) \to \bm{4}+\bm{2}\,, \quad (\bm{1},\bm{3}) \to \bm{3}\,, \quad (\bm{2},\bm{1}) \to \bm{2}\,, \quad (\bm{1},\bm{2}) \to \bm{2}\,, \quad (\bm{1},\bm{1}) \to \bm{1}\,.
\end{equation}
Meanwhile, three of the $\asu(2)\oplus\asu(2)$ gauge bosons, those corresponding to the generators
\begin{equation}
T^a_{(4)} - T^a_{(1)}\,,
\end{equation}
should get masses as part of this Higgsing process. The $T^a_{(4)} - T^a_{(1)}$ generators appear to be charged in the adjoint ($\bm{3}$) representation of the diagonal $\asu(2)$ algebra, so they must pair up with a hypermultiplet in the $\bm{3}$ representation of $\asu(2)$ to form massive vector multiplets. This $\bm{3}$ hypermultiplet is removed from the massless spectrum. Combining this information with \cref{eq:quinticsu2su2spec,eq:quinticbranching} leads to an ostensible massless hypermultiplet spectrum of
 \begin{equation}
 2\times\frac{1}{2}\bm{4}  + 84\times\bm{2} + 104\times\bm{1}\,,
 \end{equation}
 in exact agreement with \cref{eq:quinticsu2spec}.

We therefore see that, if we could Higgs the $\asu(2)\oplus\asu(2)$ gauge algebra down to $\asu(2)$, we would recover the expected $\asu(2)$ model with $\gaugedivclass = 5$ and two half-hypermultiplets of $\bm{4}$ matter. However, the $\asu(2)\oplus\asu(2)$ model in question, with the hypermultiplet spectrum of \cref{eq:quinticsu2su2spec}, lacks the bifundamental matter necessary to perform this Higgsing. The \conjecture\ then states that any attempt to construct this $\asu(2)$ model would lead to a model with an $\asu(2)\oplus\asu(2)$ gauge algebra. In other words, an $\asu(2)$ gauge algebra with the hypermultiplet spectrum of \cref{eq:quinticsu2spec} would automatically enhance to an $\asu(2)\oplus\asu(2)$ gauge algebra. The conjecture thus offers a physical explanation for the geometric observation that a quintic curve with two triple points must be reducible.

The geometric analysis of \cite{KleversEtAlExotic} identified other $\asu(2)$ models with $\bm{4}$ matter that have similar geometric obstructions. These models would be related to the quintic $\asu(2)$ model discussed here by applying Cremona transformations to the $\bP^2$ base of the elliptic fibration. It is natural to expect that these geometric obstructions can also be explained physically by the \conjecture. However, the $\asu(2)$ divisors in these examples have more complicated algebraic structures, making it more difficult to determine the enhanced gauge algebras. We therefore do not analyze these examples here, although they could serve as interesting tests of the \conjecture\ in future work.

\subsection{More exotic matter}
\label{sec:exotic-non-F-theory}

We now briefly consider exotic matter in representations not expected
from F-theory according to the classification of
\cite{KleversEtAlExotic}. Some examples of such matter representations
are the 3-index antisymmetric ($\bm{84}$) representation of $\SU(9)$,
the {\tiny$\ydiagram{2, 2}$} ($\bm{20}$) representation of $\SU(4)$, any
representation of $\gG_2$ other than the $\bm{7}$ and the adjoint,
etc.  In general, attempts to construct these representations lead to
codimension two $(4, 6)$ loci, which are associated with
superconformal sectors \cite{Seiberg:1996qx, Heckman:2013pva,
DelZotto:2014hpa}.  While there may be a physical interpretation of
these SCFT matter structures, these kinds of matter/SCFT structures
are as yet poorly understood in F-theory.  We may, however, see the
appearance of a $(4, 6)$ codimension two locus as another kind of
``enhancement'' that is forced when certain combinations of gauge
group and matter fields are tuned in an F-theory model.  In
particular, this kind of enhancement, or an inconsistency of the
theory, seems to arise whenever a theory is tuned in F-theory that
could possibly be consistent with exotic matter outside the list of
representations identified in \cite{KleversEtAlExotic}.  One class of
situations in which this $(4, 6)$ enhancement has been studied is for
the 3-index antisymmetric $\SU(9)$ representation; while there are
apparently anomaly-free 6D supergravity models that contain this
representation, any attempt to construct them in F-theory appears to
lead to a codimension-two $(4, 6)$ locus associated with the desired
matter representation \cite{AndersonGrayRaghuramTaylorMiT}.

Here we briefly analyze another simple set of 6D supergravity theories
with exotic matter that satisfy all known consistency conditions but
do not seem to have an F-theory realization.  These are theories with
gauge group $\gG_2$, no tensor multiplets, and matter in the
$\bm{27}$-dimensional exotic matter representation.  Some simple 6D
supergravity models with this representation that seem to be
anomaly-consistent are, for example, those with $T = 0$, gauge group
$\gG_2$, and a matter spectrum of $31 \times \bm{7}+ 0 \times \bm{14}
+ x \times \bm{27}$, where $\bm{14}$ is the adjoint
representation and $x = 1$ or $x = 2$.  We argue here that any attempt
to tune such a theory leads to an ``enhancement'' to a codimension two
$(4, 6)$ sector.

It is fairly straightforward to see that tuning any kind of $\gG_2$
gauge factor with matter in the $\bm{27}$ representation leads to
some kind of complications in F-theory.  As mentioned briefly earlier,
one simple way of classifying exotic matter fields is through their
contribution to the arithmetic genus of the curve supporting them in
an F-theory model, which can be computed directly from the
contribution to \cref{eq:genus} from that particular representation
in the anomaly equations \cite{KumarParkTaylorT0}. The $\bm{27}$ of
$\gG_2$ has a genus contribution of $g = 3$.  Thus, we expect only to
see this representation when $\gG_2$ is tuned on a curve with a triple
self-intersection point or other singularity of arithmetic genus 3.
At such a point, however, $(f, g)$ must vanish to orders $(6, 9)$
since the orders needed for $\gG_2$ in the Kodaira classification are
$(2, 3)$.  We thus have an automatic $(4, 6)$ point in any F-theory
model when we try to tune this representation.  The meaning of this
kind of singularity is not quite clear at this time.  In general, as
mentioned above, we
expect a superconformal field theory (SCFT) to be coupled in at such
codimension $(4, 6)$ points.  It is possible that such an SCFT
could play the role of the $\bm{27}$ matter, although it is not clear
how to make complete sense of this interpretation.  It has also been
suggested \cite{CveticHeckmanLinExoticDiscrete} that T-brane
structures may somehow enable exotic matter at codimension-two $(4,
6)$ loci in such a way that there is no SCFT coupled into the theory, and
in principle such a mechanism may operate here, but this is also poorly
understood.  While the base can be changed by blowing up the $(4, 6)$
point associated with the exotic matter, giving another base with $T
\to T + 1$, in this case the $\gG_2$ factor would generally lose
its self-intersection at the locus supporting the desired $\bm{27}$
and the matter associated with that point would be lost.
In any case, the enhancement of the Weierstrass model to a codimension
two $(4, 6)$ point may represent some kind of enhancement related to
those described in the bulk of this paper, though it is not covered by
the \conjecture\ as stated in \cref{sec:conjecture}.  It would be interesting to try to
incorporate this kind of swampland model into a larger framework
including the other cases described here that are better understood.

Another interesting class of enhancements related to some kind of
strongly coupled SCFT matter arises in the case of a tuned $\gE_8$ gauge
group.\footnote{Thanks to Yinan Wang for helpful discussions on this point.}  Again here there is a necessary enhancement to a local $(4,
6)$ point when we tune, e.g.,an $\gE_8$ factor on a curve on $\bP^2$,
associated with some kind of SCFT.   Deformations of the Weierstrass
model away from this locus, as for the other tuned $(4, 6)$ point
SCFTs presumably correspond to some kind of Higgsing that breaks the
$\gE_8$.  Some investigation of this kind of structure was carried out
in \cite{Tian:2018icz}, but it would be nice to better understand these kinds of
SCFT enhancements and their role in the \conjecture.

\subsection{Discrete gauge groups}
\label{sec:discrete}

Automatic enhancement can occur for models with discrete gauge symmetries as well, even when the desired charged matter spectrum satisfies the low-energy anomaly conditions of \cite{MonnierMooreGS}. We do not give a complete description of automatic enhancement for discrete gauge groups here. Instead, we present examples of F-theory models where a $\Z_2$ or $\Z_3$ symmetry automatically enhances to a larger group, such as $\U(1)$.

 For instance, we can consider a
$\Z_2$ model with zero tensor multiplets and 108 charged
hypermultiplets. One might imagine trying to obtain this model by
Higgsing a $\U(1)$ model with anomaly coefficient $\tilde{b}=6H$ and
108 hypermultiplets of $\chargedsinglet{1}$ matter, but this $\U(1)$ model
does not have the requisite $\chargedsinglet{2}$ hypermultiplets to perform the
Higgsing. The \conjecture\ would therefore suggest that the
$\Z_2$ gauge symmetry should automatically enhance to a
$\U(1)$. To see if this enhancement occurs explicitly in F-theory, we
can consider the $\Z_2$ Weierstrass model from
\cite{Braun:2014oya, MorrisonTaylorMultisection}, which is given by
\begin{equation}
y^2 = x^3 + \left(e_1 e_3 - \frac{1}{3} e_2^2 - 4 e_0 e_4\right)x + \left(-e_0 e_3^2 - \frac{1}{3}e_1 e_2 e_3 - \frac{2}{27}e_{2}^3 + \frac{8}{3}e_0 e_2 e_4 - e_1^2 e_4\right)\,.
\end{equation}
We let the base be $\bP^2$ to ensure there are zero tensor
multiplets, and to obtain the correct number of charged
hypermultiplets, we should let $\divclass{e_i} =
(12-3i)H$.\footnote{We could alternatively let $\divclass{e_i} = 3iH$,
but because we can exchange $e_i\leftrightarrow e_{4-i}$, the
analysis is the same for this case.} Since $\divclass{e_4}$ is
trivial, $e_4$ can be freely taken to be a perfect square, implying
that the gauge group is $\U(1)$ instead of $\Z_2$
\cite{MorrisonTaylorMultisection}. We therefore see the expected
$\Z_2 \to \U(1)$ automatic enhancement.

One can also consider the $F_1$ toric hypersurface model discussed in \cite{KleversEtAlToric}, which has gauge group $\Z_3$. Considering this model tuned over the base $\bP^2$, the Jacobian rank method of \cite{RaghuramTaylorTurnerSM} seems to indicate that this is the universal model for the gauge group $\Z_3$. However, for most values of the parameters $\cS_7, \cS_9$, the gauge group enhances, due to either the ineffectiveness or triviality of classes of some parameters in the Weierstrass model (in the latter cases, the model develops extra $\U(1)$ gauge factors because the associated parameters are automatically perfect squares, as in many of the other cases we have discussed). For example, there is an anomaly-consistent $\Z_3$ model with $(\cS_7, \cS_9) = (0, 1)$ containing 168 charged hypermultiplets (note that charges $\pm1$ and $\pm2$ are equivalent in a 6D $\Z_3$ model). However, there is an associated $\U(1)$ model (the $F_3$ toric hypersurface model of \cite{KleversEtAlToric}) with $(\cS_7, \cS_9) = (0, 1)$ that would Higgs to this $\Z_3$ model, except that for this choice of classes the $\U(1)$ model has no $\chargedsinglet{3}$ matter with which to perform the Higgsing. Thus, according to the \conjecture, we expect that the $\Z_3$ model automatically enhances to a $\U(1)$ model for this choice of $\cS_7, \cS_9$, and indeed this is the case. It appears to be true that all of the enhanced models over the base $\bP^2$ have insufficient matter to supersymmetrically Higgs to the associated $\Z_3$ model, consistent with the \conjecture. Note that there are also choices of $\cS_7, \cS_9$, such as $(\cS_7, \cS_9) = (1, 9)$ that are a priori consistent with anomalies for which the discriminant of the $F_1$ model vanishes identically, indicating that these models are actually inconsistent altogether.

It would be interesting to perform a more thorough investigation of automatic
enhancements of discrete gauge symmetries in future work.

\section{Automatic enhancement in lower-dimensional theories}
\label{sec:lower-dimensions}

\subsection{Automatic enhancement in 4D $\cN = 1$ F-theory models}

While we have focused in this paper on 6D F-theory models, at least at
the level of geometry similar enhancements are forced through tuning
Weierstrass models with certain gauge groups on sufficiently large
divisors in 4D F-theory models as well.  Because the actual gauge
group can be affected by fluxes and the superpotential, the physical
consequences of this geometric enhancement are less clear in 4D
theories, but the analogous phenomenon seems worth exploring.

As a simple example, consider an F-theory model on the base $\bP^3$,
with an $\asu(2)$ algebra tuned on the divisor $D = b H$, with $H$ the
hyperplane divisor.  The same logic as that described in
\cref{sec:su2-enhancement} applies.  Here $K_B = -4 H$, and so if $12
<b <16$, there will be an additional $\asu(2)$ algebra arising on the
divisor $f_1 = 0$, with $[f_1] = (16 - b) H$.  As in the 6D context,
the resulting global gauge group, at least at the level of geometry,
will be $(\SU(2) \times\SU(2))/\Z_2$.

\subsection{Automatic enhancement in 5D and 4D theories with 8 supercharges}

It is interesting to speculate whether some similar mechanism may
arise in 5D $\cN = 1$ and 4D $\cN = 2$ theories.  In
particular by compactifying at least some of the cases where 6D
theories have automatic enhancement on circles, it seems we can get
some lower-dimensional theories with additional gauge factors that cannot
be  completely Higgsed away.

As a simple example, consider the 6D $T = 0$ theory with an $\asu(2)$
tuned on a curve with $b = 10$ or $b = 11$, which as discussed in
\cref{sec:su2su2} is automatically enhanced to include an extra
$\asu(2)$ factor on a complementary curve with $[f_1] = 2, 1$
respectively.  If we compactify this 6D theory down to a 5D or 4D
theory with 8 supercharges on a circle $S^1$ or torus $T^2$, the
lower-dimensional theory again has a gauge group
$(\SU(2) \times \SU(2)) / \Z_2$.  The vector multiplets now include
adjoint scalars for the $\asu(2)$ factors, so there is a Coulomb
branch where, for example, the first $\asu(2)$ factor remains nonabelian
while
the second factor is broken to the $\U(1)$ subgroup so that the global
group becomes $(\SU(2) \times \U(1)) / \Z_2$.  In the 5D theory there are
no fields charged purely under the $\U(1)$ factor, so
it seems that we may have a similar situation where the containing
$G'$ theory cannot be broken down to a smaller theory with only a
gauge algebra $\asu(2)$ and the same charged matter content under that
factor.  The story is much less clear here, however, since we do not
have an analogous construction to the F-theory geometry where we can
see the forcing of the extra factor through the Weierstrass model.  If
there is some constraint analogous to the automatic enhancement
mechanism for 5D theories, for example for 5D models that can be
realized by compactifying M-theory on a Calabi--Yau threefold, this
would seem to entail some constraint on the geometry of general CY
threefolds that might hold even beyond the elliptic CY3 framework.  We
leave this question for further investigation.
It would also be interesting to look into the parallel questions for
4D models coming from compactification on a further circle.

\section{Conclusions and further directions}
\label{sec:further}

In this paper we have identified a range of circumstances in which
tuning a certain gauge group $G$ and matter content $M$ in F-theory
automatically gives rise to a larger gauge group $G'$ and matter
content $M'$.  We have analyzed a variety of constructions for
specific gauge groups $G$ that satisfy the conditions of the
\conjecture\ stated in \cref{sec:conjecture} that there is a
containing theory $G', M'$ such that the larger theory cannot be
broken down to the smaller theory with gauge group $G$ by a
supersymmetry-preserving Higgs deformation.  For each of the specific
gauge groups $G$ with generic matter content, we have used explicit
constructions of universal Weierstrass models to identify positivity
conditions and associated local conditions in terms of anomaly
coefficients that are sufficient to show that certain gauge group and
matter combinations cannot be realized in F-theory without some
additional enhanced gauge factors. These gauge group-specific
formulations of the conjecture are stronger than the general statement
since they also incorporate classes of theories that are rendered
invalid by the enhancement.

While the analysis here has focused on F-theory constructions, where
we can explicitly prove that automatic enhancement must take
place at least through standard F-theory constructions, it is interesting to
speculate that the \conjecture, and its group-specific stronger versions,
may hold more generally for 6D
supergravity theories.  Indeed, many of the known 6D supergravity
models that seem to lie in the swampland \cite{VafaSwamp} of
theories that appear consistent but have no known string realization,
and which do not violate other known F-theory constraints reviewed in
\cref{sec:constraints}, are in the class of theories identified here
that are automatically enhanced to a larger gauge symmetry when
realized in F-theory through standard constructions, or are rendered
invalid when this enhancement is not possible.  Furthermore, as
described in
\cite{MorrisonTaylorCompleteness} and reviewed in \cref{sec:su2-quotient}, another class of apparent
swampland theories, in which the massless matter does not span the
full charge lattice of the gauge group, appears to be similarly ruled
out in F-theory through the appearance of additional sections.  It would be interesting
to try to find some way of unifying the set of swampland theories that
undergo automatic enhancement to either a larger group or an invalid
F-theory model, along with those theories that automatically acquire
extra Mordell--Weil  sections in F-theory, through some
condition on the low-energy theory, which would provide a unified
swampland hypothesis for all three of these sets of theories in the
context of 6D supergravity.  Such a formulation seems a bit subtle,
however; for example, in the context of the $\asu(2) \to
(\SU(2) \times\SU(2))/\Z_2$ described in \cref{sec:su2su2}, automatic
enhancement to the extra $\asu(2)$ factor is forced on the locus $f_1
= 0$.  From the F-theory point of view, this is one particular curve
in a moduli space of curves in a given homology class, which is picked
out by the structure of the Weierstrass model.  In the 6D supergravity
theory, that homology class corresponds to a particular charge in the
string charge lattice, but the degrees of freedom associated with the
specific choice of curve $f_1$ are not clearly visible in the
low-energy theory in any way that we are familiar with; thus, in
particular, world-volume methods on a string such as those used in
\cite{KimShiuVafaBranes, LeeWeigandSwampland} do not seem sensitive to
the specific locus $f_1 = 0$ as distinct from the other string
configurations coming from branes wrapped on other cycles in the same
class.  Understanding this better would be an interesting direction for
further investigation.

Another question of interest is whether the \conjecture\ can be
rigorously proven even in the context of F-theory.  For the first case
we have discussed, that of non-Higgsable clusters, such a proof exists.
In this situation, where there is a curve $C$ of negative
self-intersection $C \cdot C \le -3$, the curve $C$ is rigid and
hence unique, and the arguments of \cite{MorrisonTaylorClusters} show
that any F-theory model with such a curve must have a non-Higgsable
gauge group of at least $\SU(3)$.  For the other cases we have
described here, however, the argument is not quite so complete.
We have shown that standard F-theory constructions with
various gauge groups and matter content $G, M$ are automatically
enhanced to $G', M'$, however, and while
we have not ruled out the possibility of exotic
F-theory constructions such as those using non-UFD structures on
singular curves \cite{KleversEtAlExotic} that could in principle
realize $G, M$ theories without enhancement, the results of
\cref{app:u1u1test} suggest that this approach will not lead to
theories that violate the conjecture.   Another way to prove the
\conjecture\ for many of the classes of examples we have described
here would be to show that the universal moduli space of theories with
at least a given
gauge group and matter content $G, M$
must be a connected set
(within the context of a class
of theories with a given F-theory base $B$, corresponding to a fixed
string charge lattice and positivity cone in the low-energy theory, and
assuming fixed anomaly coefficients for the factors of $G$).
If this is true, then every theory with $G, M$ could be reached from
every other, so that the enhanced theories $G', M'$ would need to be
connected to any smaller theory such as one with only the gauge group
and matter content $G, M$ and no enhancement, which is not allowed by
the conditions of the \conjecture.  If such an argument could be
realized, one could imagine a similar argument might apply in the more
general context of 6D supergravity without reference to
an F-theory completion, if it were possible to show that for a given string charge
lattice and positivity cone at a fixed number of tensor multiplets $T$
the moduli space of allowed supergravity theories with  at least a
given $G, M$ is connected, for fixed values of the anomaly coefficients.

\acknowledgments

We would like to thank Lara Anderson, James Gray, Daniel Harlow, Sheldon Katz, David Morrison,
Paul Oehlmann, Houri Tarazi, Cumrun Vafa, and Yinan Wang for helpful discussions.  The work of APT
and WT was supported by DOE grant DE-SC00012567. APT is supported in part by the DOE (HEP) Award DE-SC0013528. NR is supported by NSF grant PHY-1720321.

\appendix

\section{Explicit analysis of rational sections}
\label{app:ratsections}

\subsection{$(\SU(2)\times\U(1))/\Z_2$ model}
For the $\asu(2)\to (\SU(2)\times\U(1))/\Z_2$ enhancement of \cref{sec:su2u1}, we take the Weierstrass model in \cref{eq:su2} and consider situations where the divisor class $\divclass{g_2}$ is trivial. The parameter $g_2$ is then essentially a constant, and we can freely set it to a perfect square $\gamma^2$ without issue, where $\divclass{\gamma}$ is trivial. This is \emph{not} a tuning of the model when $\divclass{g_2}=0$ but is instead a simple rewriting of $g_2$. After this substitution, the Weierstrass model is described by
\begin{equation}
f = -\frac{1}{48}\phi^2 + \sigma f_{1}\,, \quad g = \frac{1}{864}\phi^3-\frac{1}{12}\phi \sigma f_{1} +\sigma^2 \gamma^2\,.
\end{equation}
This Weierstrass model is in fact in the Morrison--Park form \cite{MorrisonParkU1}\footnote{To avoid confusion with the anomaly coefficients, we use the symbol $\mpb$ to denote the parameter typically called $b$.}
\begin{equation}
\label{eq:Morrison--Park}
f= c_1 c_3 - \mpb^2 c_0 - \frac{1}{3}c_2^2\,, \quad g= c_0 c_3^2 -\frac{1}{3}
c_1 c_2 c_3 +\frac{2}{27}c_2^3 -\frac{2}{3}\mpb^2 c_0 c_2
\end{equation}
with
\begin{equation}
\mpb= 1\,, \quad c_0 = \frac{\phi^2}{64}- f_1\sigma\,, \quad c_1= -2 \gamma\sigma\,, \quad c_2 = -\frac{1}{8}\phi\,, \quad c_{3} = 0\,.
\end{equation}
It therefore admits an infinite-order generating section
\begin{equation}\label{eq:ratsecu1}
[\secx:\secy:\secz] = \left[\frac{\phi}{12}:\gamma \sigma: 1\right]\,,
\end{equation}
and the gauge algebra is $\asu(2)\oplus\au(1)$. Because
$\divclass{\gamma}$ is trivial, the only   $\au(1)$-charged matter
occurs at $\locus{\sigma=f_{1}^2 - \gamma^2 \phi}$; this locus
supports $\bm{2}_{1/2}$ matter, as can be verified with the
techniques in \cite{RaghuramTurnerOrders}.\footnote{The $\au(1)$
  charges are normalized here such that the lattice of singlet charges has
  unit spacing \cite{CveticLinU1}. While there are no charged singlets
  when $\divclass{\gamma}=0$, there are charged singlets at
  $\locus{f_1=\gamma=0}$ for other choices of $\divclass{\gamma}$. The
  charge of these would-be singlets sets the
  normalization.} This matter spectrum satisfies the anomaly cancellation conditions for $\height = -2\canonclass - \frac{1}{2}\divclass{\sigma}$, and the gauge group is $(\SU(2)\times\U(1))/\Z_2$.

\subsection{$ (\SU(2)\times\SU(2))/\Z_2$ and $\SU(2)/\Z_2$ models}
For the situations encountered in \cref{sec:su2su2,sec:su2-quotient}, we take the Weierstrass model given by \cref{eq:su2} and consider situations where the divisor class $\divclass{g_2}$ is ineffective. The parameter $g_2$ is therefore 0, leading to a Weierstrass model with
\begin{equation}
f = -\frac{1}{48}\phi^2 + \sigma f_{1}\,, \quad g = \frac{1}{864}\phi^3-\frac{1}{12}\phi \sigma f_{1}\,.
\end{equation}
This Weierstrass model admits a rational section of the form
\begin{equation}
[\secx:\secy:\secz] = \left[\frac{\phi}{12}:0:1\right]\,,
\end{equation}
which is a torsional section of order 2. The section is directly
correlated with the $\Z_2$ quotient seen in the gauge groups \cite{MorrisonTaylorCompleteness}. In fact, this torsional section can be viewed as the $\gamma \to 0$ enhancement of the rational section in \cref{eq:ratsecu1} along the lines of \cite{BaumeEtAlTorsion}.

\section{$\U(1)\times\U(1)$ model}
\label{app:u1u1model}

In \cref{sec:abelian}, we make use of a $\U(1)\times\U(1)$ F-theory
model based on the construction in
\cite{CveticKleversPiraguaMultU1}. (A more general class of F-theory
models with group $\U(1) \times\U(1)$ was constructed in \cite{CveticKPT},
but the more general form is not needed here.) The base of the elliptic fibration
is $\bP^2$, such that the resulting 6D spectrum has zero tensor
multiplets. We take the divisor classes of the parameters to be the
following:\footnote{Even though the variables are written as $s_{i}$
  in \cite{CveticKleversPiraguaMultU1}, we write them here as $S_{i}$
  to avoid confusion with variables used elsewhere in this paper.}
\begin{equation}
    \begin{aligned}
        \divclass{S_1}&=3H\,, & \divclass{S_2}&= 6H\,, & \divclass{S_3}&=9H\,,  & \divclass{S_5}&= 0H\,,\\
    \divclass{S_6}&= 3H\,, & \divclass{S_7}&= 6H\,, & \divclass{S_8}&=-3H\,, & \divclass{S_9}&= 0H\,.
    \end{aligned}
\end{equation}
The Weierstrass construction for this $\U(1)\times\U(1)$ model, with $S_8$ set to 0 due to its ineffective divisor class, is described by
\begin{equation}
\begin{aligned}
    f &= -\frac{1}{48}\left(S_6^2-4 S_5 S_7\right)^2 \\
    &\qquad\qquad + \frac{1}{6} S_9 \left(6 S_1 S_3 S_9-3 S_1 S_6 S_7-2 S_2^2 S_9+2 S_2 S_5 S_7+S_2 S_6^2-3 S_3 S_5 S_6\right)
\end{aligned}
\end{equation}
and
\begin{equation}
\begin{aligned}
    g &= \frac{1}{864}\left(S_6^2-4 S_5 S_7\right)^3+\frac{1}{27}S_2 S_9^3 \left(9 S_1 S_3-2 S_2^2\right) \\
    &\qquad\qquad - \frac{1}{72} S_9 \left(S_6^2-4 S_5 S_7\right) \left(-3 S_1 S_6 S_7+S_2 \left(2 S_5 S_7+S_6^2\right)-3 S_3 S_5 S_6\right) \\
    &\qquad\qquad + \frac{1}{36} S_9^2 \Bigg[9 S_1^2 S_7^2-3 S_3 \left(S_1 \left(2 S_5 S_7+S_6^2\right)+2 S_2 S_5 S_6\right) \\
    &\qquad\qquad\qquad\qquad\qquad -6 S_1 S_2 S_6 S_7+2 S_2^2 \left(2 S_5 S_7+S_6^2\right)+9 S_3^2 S_5^2\Bigg]\,.
\end{aligned}
\end{equation}
The discriminant is then proportional to $S_9^2$, but because $[S_9]$ is trivial, this factor does not signal the appearance of a nonabelian gauge algebra. There are two generating sections for this elliptic fibration. The first, $\ratsec{s}_Q$, is given by
\begin{equation}
[\secx_{Q}:\secy_{Q}:\secz_{Q}] =\left[\frac{1}{12} \left(S_6^2-4 (S_2 S_9+S_5 S_7)\right):-\frac{1}{2} S_9 (S_3 S_5-S_1 S_7):1\right]
\end{equation}
The second, $\ratsec{s}_R$, is given by
\begin{equation}
    [\secx_{R}:\secy_{R}:\secz_{R}] = \left[\frac{1}{12} \left(S_ 6^2-4 S_ 2 S_ 9+8 S_ 5 S_ 7\right):\frac{1}{2} (S_ 5 S_ 6 S_ 7-S_ 9 (S_ 1 S_ 7+S_ 3 S_ 5)):1\right]\,.
\end{equation}
The presence of these two independent sections implies that the gauge group is $\U(1)\times\U(1)$.

According to the original matter analysis in \cite{CveticKleversPiraguaMultU1}, the locus $\locus{S_8=S_9=0}$ supports $\twochargedsinglet{-1}{-2}$ matter, and the locus $\locus{S_7=S_9=0}$ supports $\twochargedsinglet{0}{2}$ matter. Because $[S_9]$ is trivial, the model does not support any hypermultiplets of $\twochargedsinglet{-1}{-2}$ or $\twochargedsinglet{0}{2}$ matter. There is also $\twochargedsinglet{1}{1}$ matter at
\begin{equation}
    \locus{S_7 S_8^2-S_6 S_8 S_9+S_5 S_9^2 = S_3 S_8^2 - S_2 S_8 S_9 + S_1 S_9^2=0}
\end{equation}
in the original model. The class of $S_7 S_8^2-S_6 S_8 S_9+S_5 S_9^2$ is trivial in this example, and there are no hypermultiplets of $\twochargedsinglet{1}{1}$ matter. However, $\twochargedsinglet{1}{-1}$ matter is supported at $\locus{S_3=S_7=0}$, implying that this model has 54 hypermultiplets of $\twochargedsinglet{1}{-1}$ matter.  By examining the loci where $\secy$ and $3\secx^2+f\secz^4$ for the two sections simultaneously vanish, one can verify that there are 54 hypermultiplets of $\twochargedsinglet{1}{0}$ matter and 54 hypermultiplets of $\twochargedsinglet{0}{1}$ matter.  In summary, the total spectrum is
\begin{equation}
    54 \times \twochargedsinglet{1}{-1} + 54\times  \twochargedsinglet{1}{0} + 54\times \twochargedsinglet{0}{1}\,.
\end{equation}
The anomaly coefficient for the model is given by
\begin{equation}
    \height = \begin{pmatrix} 6H&  -3H \\ -3H & 6H  \end{pmatrix}\,.
\end{equation}

It is also useful to write this model in an alternative form where the fiber is a cubic curve in a $\bP^2$ ambient space. If the coordinates for this $\bP^2$ ambient space are $[u:v:w]$, the model is described by
\begin{equation}
\label{eq:u1u1p2desc}
u\left(S_1 u^2 + S_2 u v + S_3 v^2 + S_5 u w + S_6 v w\right) + v w \left(S_7 v + S_9 w\right) = 0\,.
\end{equation}
The zero section is given by $\locus{u=v=0}$, while the two generating sections are given by $\locus{u=w=0}$ and $\locus{u=S_7 v + S_9 w = 0}$.

\section{Absence of adjoint matter on $C$ when $C + K_B$ is not effective}
\label{app:adjoint}

In various circumstances we consider nonabelian gauge factors with
anomaly coefficient $b$ such that $b + a$ is outside the positivity
cone.  In such circumstances we assert that in any F-theory
realization the nonabelian gauge factor in question cannot have
adjoint matter.  For example, this comes up in \cref{sec:su2su2},
where $[g_2]= -4 K_B - C$ is not in the cone of effective curves, but
$C' =[f_1] = -3 K_B - C$ is effective, when an $\asu(2)$ algebra is tuned
on a curve of class $C$, leading to an enhancement with at least an
$\asu(2)$ on a curve of class $C'$.  In this Appendix we give a simple F-theory based physics
argument for why  a nonabelian gauge factor on a curve of class $C'$ cannot have
adjoint matter when $C' + K_B$ is not effective, and we summarize a simple
argument for this conclusion from algebraic geometry.  We also
illustrate how
this works explicitly in a general class of situations
at $T \le 9$, which highlights the subtleties that arise in the
classification of F-theory constructions when
$T \ge 9$.

First, we give a simple physics argument based on the Morrison--Park
model \labelcref{eq:Morrison--Park}.  Given a curve class $C'= [c_3]$, we can perform a Tate tuning of an $\asu(2)$ algebra over
$c_3$, which essentially gives the Morrison--Park model with $\mpb= 0$.
When $[\mpb] = C'+ K_B$ is effective, we can deform the $\asu(2)$
model to the general Morrison--Park form by turning on a nonzero value
for $\mpb$.  This corresponds physically to breaking $\asu(2)
\to \au(1)$ by Higgsing on an adjoint field.
When $[\mpb]$ is not effective there is no such deformation, and so no
adjoint fields can be present in the original $\asu(2)$ theory.
Geometrically, this means that the curve $C'$ is a rational curve, of
genus zero.

This conclusion also follows from some basic considerations of
algebraic geometry.\footnote{Thanks to Sheldon Katz for this proof.}  We assume that $C'$ is
 an irreducible
effective curve of (arithmetic) genus $g
= 1 + C' \cdot (C' + \canonclass)/2>0$
in a complex surface
$B$  that supports an elliptic Calabi--Yau
threefold of interest for F-theory.  We furthermore assume that $B$ is
rational; this leaves out the case of the Enriques surface, which is
essentially trivial for F-theory.    The Riemann-Roch theorem for
surfaces states that
\begin{equation}
\chi ({\cal O} (C' + \canonclass))
=  C' \cdot (C' + \canonclass)/2+ \chi ({\cal O}_B)\,,
\label{eq:}
\end{equation}
where $\chi ({\cal O}_B)= 1$ since $B$ is rational, and ${\cal O} (C'
+ \canonclass)$ is the line bundle associated with $C' +
\canonclass$.  We thus have
\begin{equation}
\chi ({\cal O} (C' + \canonclass))=
h^2 ({\cal O} (C' + \canonclass))
-h^1 ({\cal O} (C' + \canonclass))
+h^0 ({\cal O} (C' + \canonclass))= g \,.
\label{eq:}
\end{equation}
Serre duality tells us that
\begin{equation}
h^2 ({\cal O} (C' + \canonclass))= h^0 ({\cal O} (-C')) \,,
\label{eq:}
\end{equation}
but $h^0 ({\cal O} (-C')) = 0$ since $-C'$ is not effective.  It
follows that
\begin{equation}
h^0 ({\cal O} (C' + \canonclass)) \geq g > 0 \,,
\label{eq:}
\end{equation}
so the line bundle ${\cal O}(C' + \canonclass)$ has nontrivial
sections and thus $C' + \canonclass$ is effective under the
assumptions made above.
  This implies, on the other hand, that if $C' + \canonclass$ is not
effective but $C'$ is irreducible and effective, then $C'$ has genus
$g = 0$, which is what we set out to prove.

To illustrate this result more explicitly,
we demonstrate the situation for $T
\le 9$ and curves of limited complexity (e.g.,  for curves of limited
degree on blow-ups of $\bP^2$),
and briefly described the complications that arise at
larger values of $T$ and for curves of higher complexity.
To proceed explicitly we consider the class of cases
where there is a basis of curves with intersection form $(+1, -1, -1,
\ldots, - 1)$ and we can write $K_B = (3, -1, -1, \ldots, -1)$.  Since
all bases can be found as multiple blowups of $\bP^2$ or the
Hirzebruch surfaces $\F_m, m \le 12$ \cite{GrassiMinimal}, this covers all
cases except the Hirzebruch surfaces with even $m$, which can be
treated with a similar but simpler argument.  For the del Pezzo
surfaces $dP_k$ given by blowing up $k < 9$ points in general position
on $\bP^2$, a basis for the cone of effective divisor classes is given
by $(1, 0, \ldots, 0)$, corresponding to a line on the original
$\bP^2$ that does not pass through any of the blown up points, $(0,
\ldots, 0, 1, 0, \ldots, 0)$, corresponding to the exceptional curve of
the $i$th blown up point, $(1, 0,
\ldots, 0, -1, 0, \ldots, 0, -1, 0, \ldots, 0)$ where the $-1$ entries
appear in positions $i, j$, corresponding to a line passing through
the corresponding two blowup points, $(2, \ldots)$ with up to $5$ $-1$
entries (and the remaining entries being 0), corresponding to a conic
passing through up to 5 of the blowup points, $(3, \ldots)$ with up to
$k <9$ $-1$ entries (or one $- 2$ and up to seven $- 1$s),
corresponding to irreducible cubics, etc..
For curves of degree $n > 3$, the set of possibilities can be complicated
by constraints on the possible multiplicities of singularities (see, e.g., \cite{KleversEtAlExotic}).
When the points are not in
general position, the effective cone is slightly larger and includes
curves of self-intersection $C \cdot C <-1$, corresponding to curves
of degree $d$ in the original $\bP^2$ that pass through more than $d^2
+1$ points (possibly including multiplicities).

If a curve  $C'$ is irreducible, we have $(K_B + C') \cdot C' = 2 g
-2$, so the condition that $g >0$ is that $(K_B + C') \cdot C' \ge
0$.  Writing $C' = (n,  -m_1, -m_2, \ldots, -m_T)$ in the given basis, this condition becomes
\begin{equation}
\label{eq:genus-inequality}
 n (n -3) \ge \sum_{i} m_i (m_i -1) \,.
\end{equation}
For $n = 3$, the only solutions to this condition have all $m_i = 0,
1$, and so
\begin{equation}
 C' + K_B = (n -3, 1 - m_1, \ldots, 1, m_T) = (0, 0\ {\rm or}\ 1\,,
\ldots,0\ {\rm or}\ 1)
\end{equation}
is clearly in the effective cone.
For $n = 4$, we can have up to two values of $m_i = 2$.  But we then
have, e.g., the worst case scenario (largest $m_i$s)
\begin{equation}
 C' + K_B = (1, -1, -1, 0, \ldots, 0) \,,
\end{equation}
which again lies in the effective cone.
For $n = 5$, we can have $m_i$s above 1 in the combinations $(3, 2,
2)$ or $(2, 2, 2, 2, 2)$.  These give (with all other $m$s equal to $1$)
\begin{equation}
 C' + K_B = (2, -2, -1, -1, 0, \ldots, 0), (2, -1, -1, -1, -1, -1, 0, \ldots, 0)\,,
\end{equation}
which again are easily seen to be in the effective cone.

A similar argument holds for larger values of $n$, as long as $T \le
9$, though as $n$ becomes increasingly large the set of effective
curves becomes more subtle; for example, $C' + K_B = (3, -2, -1^7)$
saturates \cref{eq:genus-inequality}, and depends on the fact that a
cubic can be found on $\bP^2$ with a double point at one chosen point
and passing through seven other arbitrary points.  When $T \ge 9$,
however, we can also see more subtle issues arising related to the
choice of which bases support an elliptic fibration, as well as what
kinds of singularities are possible on a curve of given degree.  For
example, at $T = 9, n = 6$, we can choose $C' = (6, -2, \ldots, - 2)$.
We then have $C' + K_B = (3, -1, \ldots, -1) = - K_B$.  We cannot,
however, blow up $\bP^2$ at 9 generic points and get an acceptable
F-theory base; the base that results from blowing up at 9 generic
points has a rigid cubic in the class $K_B$, so the Weierstrass model
has $f, g$ that generically vanish to orders $(4, 6)$ on $K_B$.  For
$T = 9$, this still clearly works as $K_B = (3, -1, \ldots, -1)$ is
effective (and non-rigid) on any acceptable F-theory base.  The
situation, however, clearly becomes more complicated for larger $T$,
since choosing curves of the form $C' = (6, \ldots)$ gives classes $C'
+ K_B = (3, \ldots)$ with 9 $-1$ entries, curves of the form $C' = (7,
\ldots)$ can have $m$s of, e.g., $14 \times 2$, etc..  Some of the
complications that arise in determining the generators of the
effective cone at larger $T$ are discussed further in,
e.g., \cite{TaylorWangNon-toric}.  An argument of this form must in
principle work for all $T$ and all allowed F-theory bases, since the
more abstract mathematical argument above makes the statement rigorous
at all $T$.

\section{Another test of automatic enhancement in a $\U(1)$ model}
\label{app:u1u1test}
In this appendix, we perform an additional test of automatic enhancement for the $r=s=1$ model in the $T=1$ infinite family of \cref{eq:chargefamilyspectrumT1}. Recall that the charged hypermultiplet spectrum for this model is
\begin{equation}
    \label{eq:r1s1u1spectrum}
    96\times\chargedsinglet{1} + 48\times\chargedsinglet{2}\,,
\end{equation}
and we expect an automatic enhancement to a $\U(1)\times\U(1)$ model with a charged hypermultiplet spectrum of the form
\begin{equation}
    48\times\twochargedsinglet{1}{0} + 48\times\twochargedsinglet{0}{1} + 48\times\twochargedsinglet{1}{1}\,.
\end{equation}
Because there is one tensor multiplet, the base of an F-theory model realizing this spectrum would be $\F_n$. If we attempt to construct this $\U(1)$ model using the Morrison--Park tuning of \cref{eq:Morrison--Park} with an $\F_n$ base, the parameters should have the following divisor classes:
\begin{equation}
    \divclass{\mpb} = -2\canonclass\,, \quad \divclass{c_m} = -m\canonclass\,.
\end{equation}
Because $\divclass{c_0}$ is trivial, we see an enhancement to $\U(1)\times\U(1)$ as discussed in \cref{sec:abelian-infinite}.

One might imagine circumventing this enhancement by obtaining the $\U(1)$ model from constructions that typically give higher charges.\footnote{The authors thank Yinan Wang for helpful discussions on this point.} Let us focus on the charge-3 construction from \cite{Raghuram34}, which supports charge-3 matter at $\locus{\eta_{a}=\eta_{b}=0}$. It is known that the construction is equivalent to the Morrison--Park form if either $\divclass{\eta_a}$ or $\divclass{\eta_b}$ is trivial. But if we choose non-trivial $\divclass{\eta_a}$ and $\divclass{\eta_b}$ such that $\divclass{\eta_a}\cdot\divclass{\eta_b}=0$, the model would only support hypermultiplets with charges $\pm1$ and $\pm2$. This model does not seem to be in Morrison--Park form, and one might hope it would realize \cref{eq:r1s1u1spectrum} without an automatic enhancement.

However, various effects either render the model invalid or secretly put it into Morrison--Park form. The charge-3 construction contains a parameter $\overline{\phi}$ that is important for its structure. If $\overline{\phi}$, $\eta_{a}$, and $\eta_{b}$ share a common nontrivial factor, $(f,g,\Delta)$ vanish to orders $(4,6,12)$ at a codimension-one locus. Alternatively, if $\overline{\phi}$ can be written as
\begin{equation}
    \phi_{a} \eta_{a} + \phi_{b}\eta_{b}\,,
\end{equation}
we can relate the charge-3 construction to the Morrison--Park form through the following map of parameters:\footnote{This can be understood through the normalized intrinsic structure \cite{KleversEtAlExotic, Raghuram34} of the charge-3 construction.}
\begingroup
\allowdisplaybreaks
\begin{align}
    \phantom{c_3}
    &\begin{aligned}
        \mathllap{\mpb} &= b_{(2)} \eta_a^2 + 2 b_{(1)} \eta_a \eta_b + b_{(0)}\eta_{b}^2\,,
    \end{aligned} \\
    &\begin{aligned}
        \mathllap{c_3} &=-\left(t_{(0)}\eta_b^3 +3t_{(1)} \eta_a \eta_b^2 +3t_{(2)} \eta_a^2 \eta_b +t_{(3)}\eta_a^3\right)  \\
        &\quad + \frac{1}{12}\left(b_{(2)} \eta_a^2 + 2 b_{(1)} \eta_a \eta_b + b_{(0)}\eta_{b}^2\right)\left(b_{(0)} \eta_b \phi_a+b_{(1)} \eta_a \phi_a-b_{(1)} \eta_b \phi_b-b_{(2)} \eta_a \phi_b\right)\,,
    \end{aligned} \\
    &\begin{aligned}
        \mathllap{c_2} &= \frac{1}{4}\left(h_{(2)} \eta_a^2 + 2 h_{(1)} \eta_a \eta_b + h_{(0)}\eta_{b}^2\right) +\frac{1}{144}\left(b_{(1)}^2-b_{(0)}b_{(2)}\right)\left(\phi_{a} \eta_{a} + \phi_{b}\eta_{b}\right)^2 \\
        &\quad +\frac{1}{96}\left(b_{(2)} \phi_b^2 - 2 b_{(1)} \phi_a \phi_b + b_{(0)}\phi_{a}^2\right)\left(b_{(2)} \eta_a^2 + 2 b_{(1)} \eta_a \eta_b + b_{(0)}\eta_{b}^2\right)\\
        &\quad +\frac{1}{4}\left(t_{(1)}  \eta_b^2 +2t_{(2)} \eta_a \eta_b +t_{(3)}\eta_a^2\right)\phi_b - \frac{1}{4}\left(t_{(0)}  \eta_b^2 +2t_{(1)} \eta_a \eta_b +t_{(2)}\eta_a^2\right)\phi_a\,,
    \end{aligned} \\
    &\begin{aligned}
        \mathllap{c_1} &=  \frac{1}{1728}\left(b_{(2)} \phi_b^2 - 2 b_{(1)} \phi_a \phi_b + b_{(0)}\phi_{a}^2\right)\left(b_{(0)} \eta_b \phi_a+b_{(1)} \eta_a \phi_a-b_{(1)} \eta_b \phi_b-b_{(2)} \eta_a \phi_b\right) \\
        &\quad -\frac{\eta_a}{48}\left(48\lambda_{(1)}-2h_{(1)}\phi_a + 2h_{(2)}\phi_b+t_{(1)}\phi_a^2 - 2t_{(2)}\phi_a\phi_b+t_{(3)}\phi_b^2\right) \\
        &\quad -\frac{\eta_b}{48}\left(48\lambda_{(0)}-2h_{(0)}\phi_a + 2h_{(1)}\phi_b+t_{(0)}\phi_a^2 - 2t_{(1)}\phi_a\phi_b+t_{(2)}\phi_b^2\right)\,,
    \end{aligned} \\
    &\begin{aligned}
        \mathllap{c_0} &=-f_2 +\frac{1}{82944}\left(b_{(2)} \phi_b^2 - 2 b_{(1)} \phi_a \phi_b + b_{(0)}\phi_{a}^2\right)^2 \\
        &\quad + \frac{1}{576}\left(h_{(0)} \phi_a^2 - 2 h_{(1)} \phi_a \phi_b + h_{(2)}\phi_{b}^2\right) \\
        &\quad -\frac{1}{1728}\left(t_{(0)}\phi_a^3 -3t_{(1)} \phi_b \phi_a^2 +3t_{(2)} \phi_b^2 \phi_a -t_{(3)}\phi_b^3\right)-\frac{1}{12}\left(\lambda_{(0)}\phi_a - \lambda_{(1)}\phi_b\right)\,.
    \end{aligned}
\end{align}
\endgroup
This mapping also applies if $\divclass{\overline{\phi}}$ is ineffective; we simply set $\phi_a$ and $\phi_b$ to 0.

With these effects in mind, we now show that one cannot use the charge-3 construction to successfully obtain \cref{eq:r1s1u1spectrum} without an automatic enhancement to $\U(1)\times\U(1)$. We go through the possible choices of divisor classes and demonstrate that they all lead to enhancement or introduce some problem. According to the matter analysis in \cite{Raghuram34}, we should let
\begin{equation}
    \divclass{\secz}=-2\canonclass\,, \quad \divclass{\eta_a}\cdot\divclass{\eta_b} = 0
\end{equation}
to obtain the desired charged hypermultiplet spectrum. This choice implies that
\begin{equation}
    \divclass{f_2}= 0\,, \quad \divclass{\overline{\phi}} = \canonclass + \divclass{\eta_a}+\divclass{\eta_b}\,.
\end{equation}
If the model simplifies to Morrison--Park form, $\divclass{c_0}=\divclass{f_2}$ is trivial, and the divisor classes ensure we see the expected automatic enhancement. We can therefore limit our attention to situations where $\overline{\phi}$ is effective, eliminating many possibilities for $\divclass{\eta_a}$ and $\divclass{\eta_b}$. When they are both multiples of $F$, for instance, $\overline{\phi}$ is ineffective, and we recover Morrison--Park form. A similar argument shows that this same effect occurs on $\F_0$ when both $\divclass{\eta_a}$ and $\divclass{\eta_b}$ are multiples of $S$. Since the only way to satisfy $\divclass{\eta_a}\cdot\divclass{\eta_b}=0$ on $\F_0$ is to have both $\divclass{\eta_a}$ and $\divclass{\eta_b}$ be multiples of either $S$ or $F$, we know that automatic enhancement occurs for every possibility on $\F_0$, and we can focus on $\F_{n\ge 1}$.

To proceed, it is helpful to write
\begin{equation}
    \divclass{\eta_a} = \alpha\left(S+a F\right)\,, \quad \divclass{\eta_b} = \beta\left(S+b F\right)\,,
\end{equation}
where $\alpha$, $\beta$ are positive integers and $a$, $b$ are non-negative rational numbers such that $\alpha a$, $\beta b$ are integers.\footnote{If $\divclass{\eta_a}$ or $\divclass{\eta_b}$ is ineffective, the model reduces to Morrison--Park form (with a different mapping between parameters). It also supports nonabelian gauge factors that can often be Higgsed away. However, these situations do not offer a way of evading the automatic enhancement observed in Morrison--Park form. Additionally, if only one of $\divclass{\eta_a}$ or $\divclass{\eta_b}$ is proportional to $F$, $\divclass{\eta_a}\cdot\divclass{\eta_b}$ is non-zero. The case where both are proportional to $F$ was discussed above.} Here, $S$ and $F$ are the divisors on $\F_n$ such that $S\cdot S=-n$ and $F\cdot F=0$. In order to have $   \divclass{\eta_a}\cdot\divclass{\eta_b}$ be $0$, we require that
\begin{equation}
     -n + a +b = 0\,.
\end{equation}

First, consider cases where $0<a,b<n$. Then, both $\eta_a$ and $\eta_b$ are reducible, and they share a common factor $v$ with divisor class $S$. We also know that the intersection products
\begin{equation}
    \divclass{\eta_a}\cdot S = -\alpha(n-a)\,, \quad \divclass{\eta_a}\cdot S = -\beta(n-b)
\end{equation}
are negative. And since $\alpha, \beta\ge 1$, the intersection product $\divclass{\overline{\phi}}\cdot S$ is negative as well:
\begin{equation}
\begin{aligned}
    \divclass{\overline{\phi}}\cdot S  &= \left(\canonclass+\divclass{\eta_a}+\divclass{\eta_b}\right)\cdot S \\
    &= n-2 -\alpha(n-a) - \beta(n-b) \\
    &\le  n-2 -(n-a) - (n-b) = -2\,.
\end{aligned}
\end{equation}
Therefore, either $\divclass{\overline{\phi}}$ is ineffective---implying the model is in Morrison--Park form---or $\overline{\phi}$ is proportional to $v$---implying that the model has a codimension-one $(4,6)$ singular locus and is invalid.

This leaves the possibility that either $a$ or $b$ is 0. Without loss of generality, we take $b=0$ and $a=n$. The divisor classes are
\begin{equation}
    \divclass{\eta_b} = \beta S\,, \quad \divclass{\eta_a} = \alpha(S+nF)\,,
\end{equation}
and $\eta_{b}$ is essentially $v^{\beta}$ with a numerical coefficient. Additionally,
\begin{equation}
    \divclass{\overline{\phi}}\cdot S = \left(\canonclass+\divclass{\eta_a}+\divclass{\eta_b}\right)\cdot S  = n+2 - \beta n\,.
\end{equation}
Because $\beta$ is positive, this intersection number is negative, implying either that $\divclass{\overline{\phi}}$ is ineffective or that $\overline{\phi}$ is proportional to $v^{\beta}$ and hence $\eta_{b}$. Both of these possibilities ensure that the model reduces to Morrison--Park form.

To summarize, if we attempt to use the charge-3 construction to
realize the matter spectrum in \cref{eq:r1s1u1spectrum}, we either end
up with an invalid model or one in Morrison--Park form that exhibits
automatic enhancement. This analysis also applies to the charge-3
construction from \cite{KleversEtAlToric}, which a specialization of
the one considered here. Similarly, the charge-4 construction from
\cite{Raghuram34} does not offer a way of realizing this matter
spectrum that does not involve the Morrison--Park form. These results support the idea that
the automatic enhancement is a genuine physical phenomenon rather than an
artifact of the Morrison--Park tuning. Furthermore, since the charge-3
construction relies on the mathematical structure of non-UFD
Weierstrass models, this analysis suggests more generally that non-UFD
models will not in general produce models that contain only
generic charges that can be realized through UFD
constructions.

\bibliographystyle{JHEP}
\bibliography{references}

\end{document}